\documentclass[aps,prl,preprint,groupedaddress]{revtex4-1}

\usepackage{graphicx}
\usepackage{dcolumn}
\usepackage{bm}
\usepackage{color}
\usepackage{comment}

\begin{document}


\title{Information cascade on networks and phase transitions
}



\author{Masato Hisakado}
\email{hisakadom@yahoo.co.jp} 
\affiliation{
* Nomura Holdings, Inc., Otemachi 2-2-2, Chiyoda-ku, Tokyo 100-8130, Japan} 

\author{Kazuaki Nakayama}
\email{nakayama@math.shinshu.ac.jp}
\affiliation{
\dag Department of Mathematics,
Faculty of Sciences, 
Shinshu University \\
Asahi 3-1-1, Matsumoto, Nagano 390-8621, Japan}

\author{Shintaro Mori}
\email{shintaro.mori@hirosaki-u.ac.jp}
\affiliation{
\dag Department of Mathematics and Physics,
Graduate School of Science and Technology, 
Hirosaki University \\
Bunkyo-cho 3, Hirosaki, Aomori 036-8561, Japan}


\date{\today}

\begin{abstract}
Herein, we  consider a voting model for information cascades on several types of networks---a random graph,  the Barab\'{a}si--Albert(BA) model, and lattice networks---by using one parameter $\omega$; $\omega=1,0, -1$ respectively correspond to these networks. Our objective is to study the relation between the phase transitions and networks using 
 the parameter, $\omega$ which is related to  the size of hubs.
 We discuss the differences between the phases in which  the networks depend. In $\omega\ne -1$, without a lattice, the following two types of phase transitions can be observed: information cascade transition and super-normal transition. The first is the transition between a state where most voters make correct choices and a state where most of them are wrong. This is an absorption transition that belongs to the non-equilibrium transition. In the symmetric case, the phase transition is continuous and the universality class is the same as  nonlinear P\'{o}lya model. In contrast, in the asymmetric case, there is a discontinuous phase transition, where the gap  depends on the network. As $\omega$ increases, the size of the hub and the gap increase. Therefore, a network that has hubs has a greater effect through this phase transition. The critical point of information cascade transition does not depend on $\omega$. The super-normal transition is the transition of the convergence speed, and  the critical point of the convergence speed transition depends on $\omega$. At $\omega=1$, in the BA model, this transition disappears, and the range where we can observe  the phase transition is   the  same as  that in the percolation model on the network.
 Both phase transitions disappear at $\omega=-1$ in the lattice case.
 In conclusion, as the performance  near the lattice case, $\omega\sim-1$ exhibits  the best performance  of the voting in all networks. 
 As the hub  size decreases,  the performance improves.
 Finally, we show the relation between the voting model and  the elephant walk model.  

\hspace{0cm}
\vspace{1cm}
\end{abstract}


\maketitle

\bibliography{basename of .bib file}

\section{I. Introduction}

Complex networks have been studied over the past 20 years \cite{nw}.
Network hubs have been particularly examined.
Hubs can be observed in real networks.
Researchers have considered hubs to play an important role in networks.
Network analysis can be applied to multidisciplinary areas such as sociology, social psychology, ethnology, and economics. 
In statistical physics, statistical models of networks related to phase transitions are discussed.  
The study of these topics has extended the field and affects research on complex systems 
\cite{galam,Cont,Egu,Stau,Curty,nuno}.
In these areas, phase transition, which depends on the network, can be observed in several models.
In particular, the percolation model of a network is well known  because of  its phase transition \cite {nw}. 
The transition point depends on the network in the percolation model, which is discussed using the Molly and Reed network conditions \cite{nw}.
The transition point decreases as the  hub size increases.
This implies that large hubs can affect the entire network.
In this paper, we discuss how the network affects the information cascade model. 
In particular, we show the effects of hubs on the non-equilibrium phase transition.

Humans estimate public perception by observing the actions of other individuals, following which they exercise a choice similar to that of others.
This is called herding behavior.
This phenomenon is referred to as social learning or imitation. 
As it is usually sensible to do what other people are doing, collective herding behavior is assumed to be the result of a rational choice according to public perception. 
This is the correct strategy in ordinary situations.
However, this approach may lead to arbitrary or even erroneous decisions because a macro-phenomenon is known as  the information cascade \cite{Bikhchandani}.
As the performance we   consider whether  the wrong  information will be corrected or not in the process.

There is a problem with how people obtain public perception.
In our previous studies, we discussed the voting model for several networks \cite{Hisakado4,Hisakado5}.
One of these networks is a one-dimensional (1D) extended lattice.
In this case, the voting rate oscillates and there is no information cascade.
We observed information cascade transitions and super-normal transitions in  the scale-free network model and random graph.
We observed that the transition point does not depend on these networks.
In contrast, the super-normal transition is the transition of convergence speed depending on the network.
In this study, we integrate  these conclusions  using a parameter $\omega$  that changes the network continuously and examine the effects of the network---a random graph,  
 the Barab\'{a}si--Albert(BA) model, and the lattice model; $\omega=1,0, -1$ respectively correspond to these models \cite{Hisakado7}.
$\omega$ is related to the size of hubs.
As $\omega$ increases the size of hubs increases.
The purpose of this paper is to  clarify the  effects of  hubs in the information cascade.

We use  a sequential voting model \cite{Hisakado3,Hisakado2} to discuss the information cascade.
In this analysis,  we can confirm the "voter" and his /her referred "voters". 
In this model, public perception is represented by 
$r$ votes selected from the previous votes.
The relation between voters is a network.
There are two types of voters: herders and independents.
Independents vote independently and play the role of noise.
Herder behavior is known as an influence response function.
This is an important function for deciding  opinion in the network.
Empirical and experimental evidence has confirmed the assumption that individuals follow threshold rules when making decisions in the presence of social influence \cite{watts2,W2, Mori3}.
This rule posits that individuals  switch between two choices only when a sufficient number of others have adopted the choice.
We refer to individuals such as voters as digital herders.

In this model, there is a transition between a state in which most voters make the correct choices and a state in which most voters are wrong.
This is an absorption transition that belongs to a non-equilibrium transition \cite{Hin,Lu}.
In the symmetric case, we demonstrate that the phase transition is continuous and the universality class is the same as
the nonlinear P\'{o}lya model.
The network affects only the  super-normal transition point.
In this paper, we discuss the relation between a universal function and  networks.
In contrast, in the asymmetric case, it is a discontinuous phase transition.
The gap in the  discontinuous phase transition depends on the network.

 The remainder of this paper is organized as follows.
In Section II, we introduce networks with  the parameter $\omega$.
In Section III, we introduce our voting model  
 mathematically  and define
 two types of voters: independents and herders.
In Section IV, we combine the voting model with the networks and consider the phase transition.
In Section V, numerical simulations are performed to confirm the phase transition.
Finally, the conclusions are presented in Section VI. 
In Appendix A, we show the relation between our voting model and the elephant walk model \cite{ele}.

\section{II. Network}
\begin{figure}[h]
\includegraphics[width=110mm]{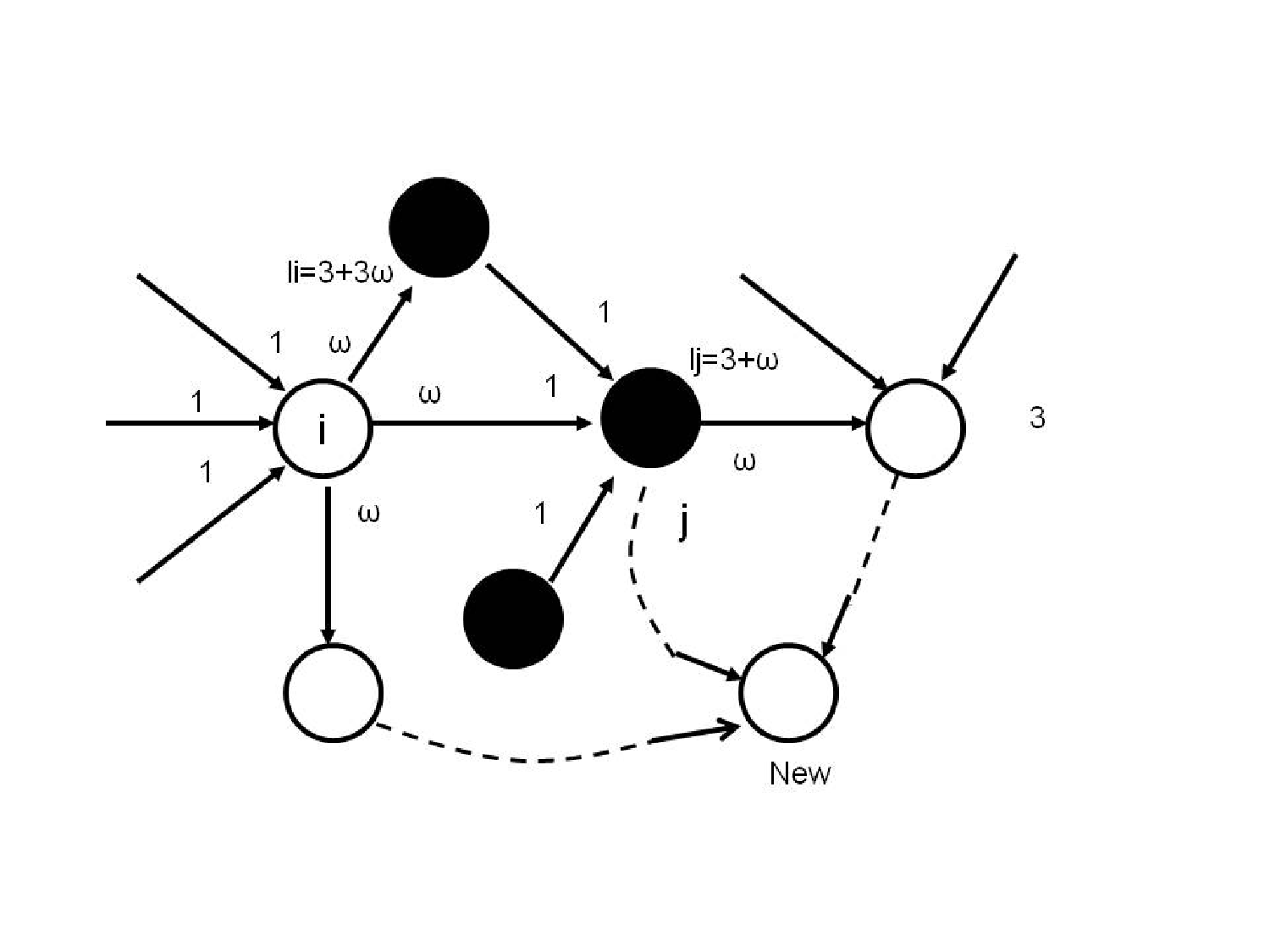}
\caption{
Sample graph with $r=3$. Voter $i$ selects three voters and is selected by three voters.
The arrow is from the selected voters to selecting voter.
The weight of the outgoing arrow is $\omega$ and
incoming arrow is 1.
The popularity is set to each voter and 
the some of the weights of incoming arrows and outgoing arrows.
Hence, the voter $j$ obtains  information from the voter $i$.  
The popularity for voter $i$   is $l_i=3+3\omega$ and that for  voter $j$ is $l_j=3+\omega$.
The color  of the node shows   the  voting of the voter.
The site is white, as he/she voted for candidate $C_0$. 
 }
\label{BA}
\end{figure}

In this section, we first define how to create voter networks according to \cite{Hisakado7}.
In the next section, we define how voters decide on their opinions.
The network means  that the voter selects referred voters to decide their opinion.
The voter corresponds to the node.
When the voter joins the network, in the first step, the voter selects $r$ voters to refer the information.
This relation corresponds to the edge.
In the second step, the voter decides his/her opinion using the information.
When the voter is   an independent, he/she does not use this information.
We consider the case wherein a voter selects $r$ voters based on popularity.
This process is sequential, as in the BA model \cite{BA}.
When voter $i$ joins at time $t=i$, site $i$
selects $r$ voters for connections, as
shown in Fig.\ref{BA}.
We show that the incoming arrows correspond to the seed popularity that all voters have when they join.
Seed popularity is $r$, the number of in-degrees.

We denote the number of incoming and outgoing edges of
node $i$ as $k_{i}^{IN}$ and $k_{i}^{OUT}$, respectively.
The degree of node $i$ is given by $k_{i}=k_{i}^{IN}+k_{i}^{OUT}$.
The popularity $l_i$ of node $i$ is defined as
\begin{equation}
l_{i}=k_{i}^{IN}+\omega \cdot k_{i}^{OUT}. \label{eq:pop}
\end{equation}
In the BA model the weight 1 for the in-degree and out-degree.
We set the weight $\omega$ for the out-degree.
The latest popularity is  used for the evolution of the network and is updated with time.
The probability that node $i<t$ is selected by node $t$ is
\begin{equation}
P(\mbox{node} \,\, i \,\, \mbox{is selected by node}\,\,t)=\frac{l_i}{\sum_{s=1}^{t-1} l_s}.
\label{eq:prob}
\end{equation}
As  popularity increases, the probability that the site is selected increases.
We show the   initial  steps in Appendix B.

We can extend this to the  negative $\omega$.
In this case, a negative feedback can be observed.
Note that if $l_i<0$, we set $l_i=0$ when $\omega<0$, in the case where $-r/\omega$ is indivisible.
Popularity decreases as the number of incoming links of the voter increases.
After the $t$-th voter joins, the total number of $l_j$ after the $t$-th voter joins is $\sum_j l_j=(\omega+1)r(t-r+1)$, where $t\geq r$.
$l_i$ corresponds to the popularity of voter $i$.
Using the parameter $\omega$, we can represent several networks using the same process.

In this network,  the voter joins the network one by one. 
This   corresponds to  sequential voting, which is explained in the next section.
Using this system  the voters can refer without fail  because the voter creates the network and decides the opinion at the same time.
When we can separate   the creation of the network and  sequential voting,   the ordering of the votes  may be the problem generally,  because the referred voter has not decided the opinion yet.
If we assume that  the voters can  refer to  the previous  votes  without fail,   the order of the voting does not affect our conclusions.

When $\omega=1$ and $\omega=0$, the network becomes a BA model \cite{BA} and  a random graph, respectively.
This is because when $\omega=0$, all voters are selected with the same probability.
In this case, the older node has many outgoing arrows because the lifetime of the older nodes is longer than that of the new node.
Hence, the probability that a node is selected 
 in  life increases with age. 
We consider the range of negative $\omega$ to be $-1\leq\omega<0$.
The maximum number of selections was $\lceil (-r/\omega) \rceil$ where $\lceil x\rceil$ is the ceiling function.
This means that there is a maximum hub size of $-1\leq\omega<0$.
There is no maximum hub size in $0\leq\omega$.
 As the size of the hub increases, $\omega$ increases owing to positive feedback.
 
\begin{figure}[h]
\includegraphics[width=110mm]{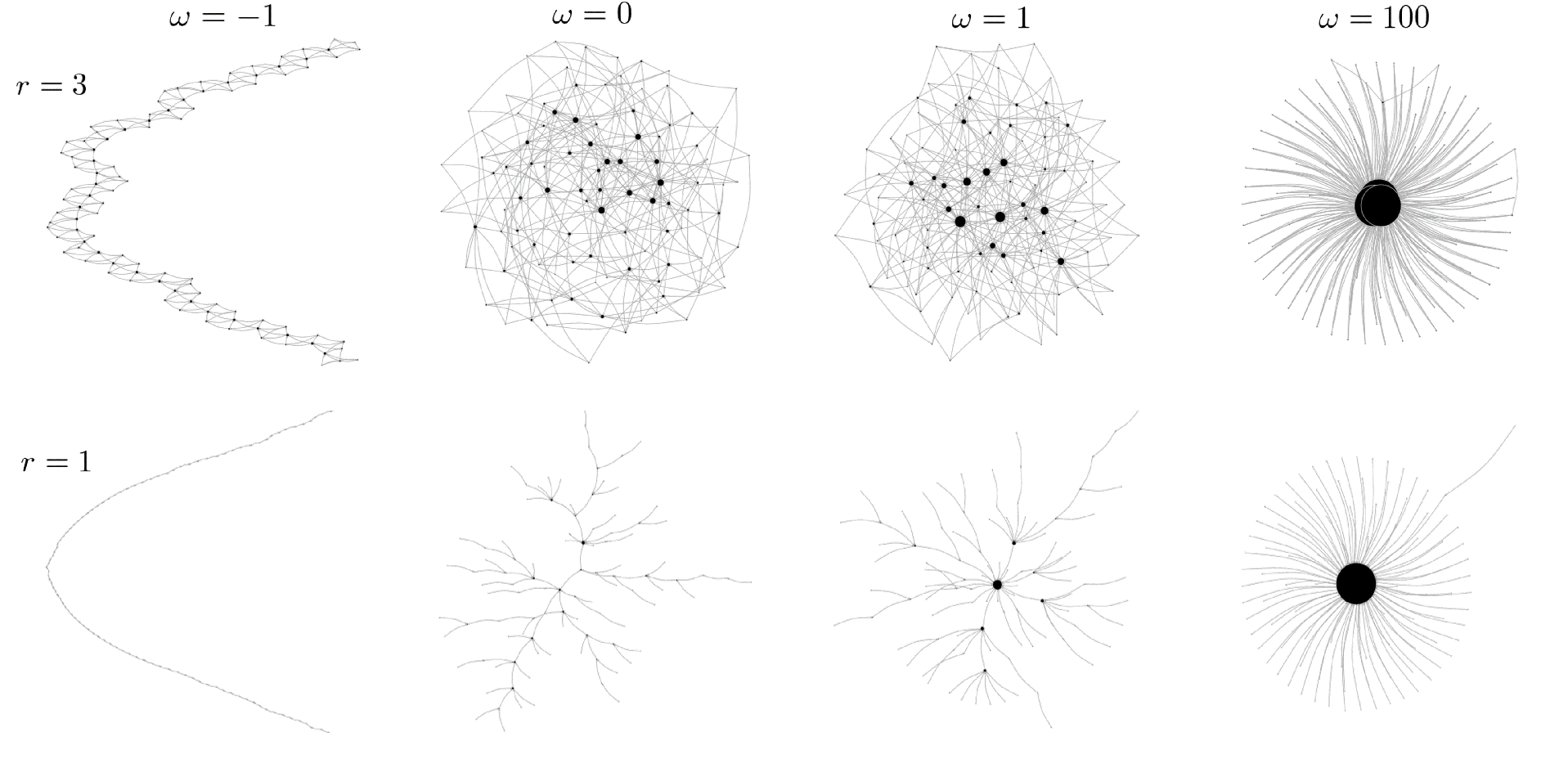}
\caption{Sample networks with $\omega \in \{-1,0,1,100\}$ in 100 steps.
 When $r=1$, the network becomes a tree network. When $r=3$,
 $\omega=1$ corresponds to the BA model. When $\omega>0$, the network is  scale-free.
 When $\omega=0$, the network is a random graph.
 When $\omega=-1$, the network corresponds to an extended lattice. }
\label{sample}
\end{figure}

Samples of the networks are shown in Fig.\ref{sample}.
The  upper side shows the case of $r=3$,
which corresponds to tree networks, and the lower side shows the case of $r=1$.
We confirmed that this method represents several types of networks.

\begin{figure}[h]
\includegraphics[width=110mm]{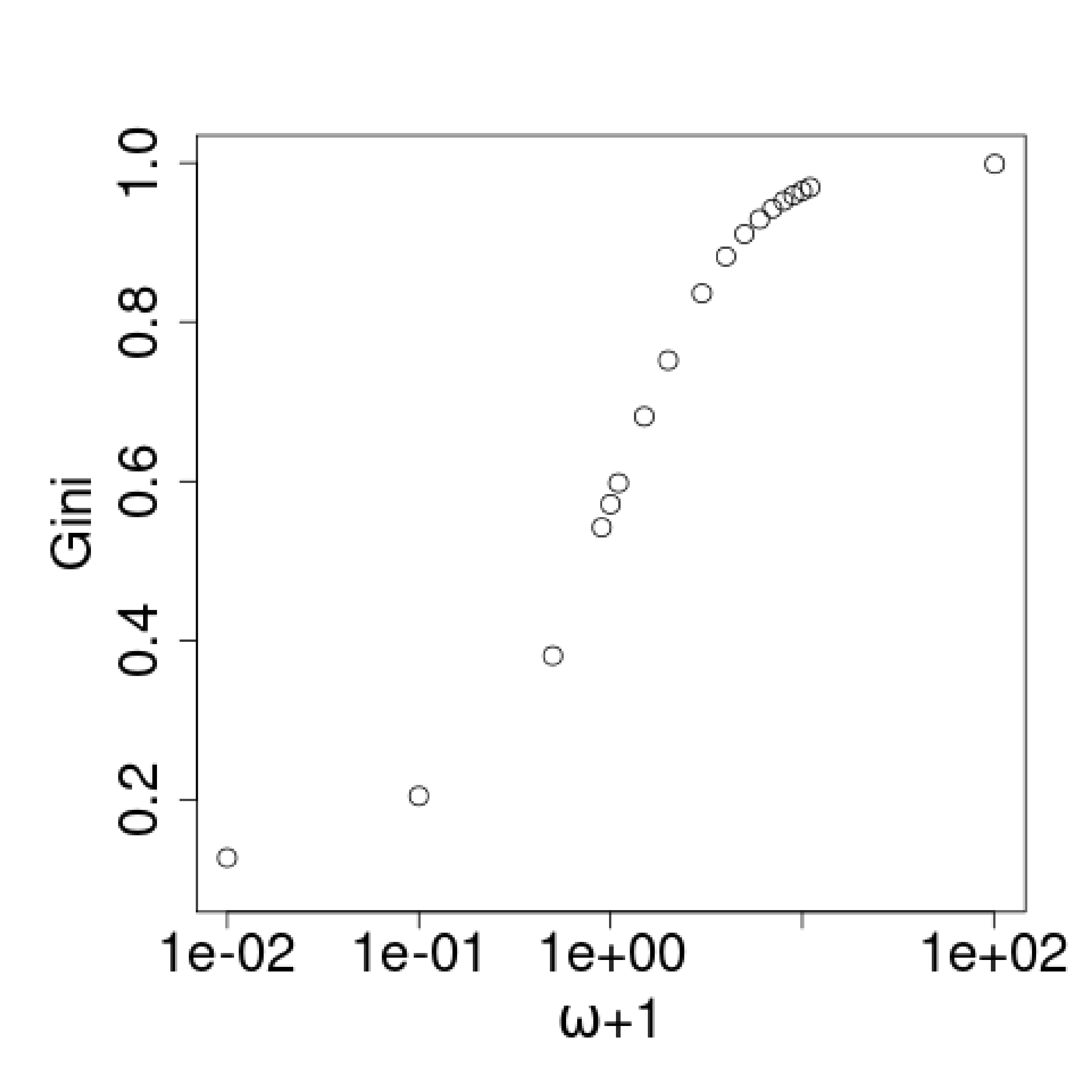}
\caption{Relation between  the Gini coefficient and $\omega$. As $\omega $ increases, most of the links would concentrate on  certain hubs. }
\label{Gini}
\end{figure}

Next, we discuss the relation between $\omega$ and the hub size.
We show that some hubs gather almost all links using  the Gini coefficient in Fig.\ref{Gini}.
This is the concentration of links to hubs.
The Gini coefficient is the index of concentration and shows a number from 0 to 1.
As the concentration increases, the Gini coefficient increases.
We can confirm that the concentration appears $\omega>0$ and the effect of hubs increases as $\omega$ increases.

\section{III. Voting Model }

In this section, we introduce the voting model.
It is the process of deciding an opinion.
We model the voting behavior of two candidates, $C_0$ and $C_1$,
at time $t$,  and $C_0$ and $C_1$ have $c_0(t)$ and $c_1(t)$ votes, respectively.
At each time step, one voter votes for one candidate, indicating that voting is sequential.
Hence, at time $t$, the $t$-th voter votes, and the total number of votes is $t$.
Voters are allowed to view them. 
$r$ previous votes for each candidate are selected as  public perception, where $r$ is a constant.
The voter  votes  simultaneously when they join the network.
Therefore,  the herder obtains the information  from the voters who have already voted. 
It is the process of  creating the network.
In the next step the voter decides his/her opinion using  information.
The  edge shows the flow of the information. The  direction is from the information provider (selected voter) to the information user (selecting voter).


In our model, we assume an infinite number of voters of each of the two types: independents and herders.
Independent votes for candidates $C_0$ and $C_1$
have probabilities $1-q$ and $q$, respectively.
Their votes are independent from others' votes; that is, 
their votes are based on their fundamental values.
We assume $q\geq 1/2$.
Independent does not refer  to the other voters but is refereed by other voters.
Hereafter, we set $C_1$ as the correct candidate, and the  voting ratio  to candidate $C_1$ is the correct ratio used to consider  voting performance.

Herders' votes are  based on the number of previous $r$ votes.
Note that the voter does not necessarily refer to the latest $r$ votes. 
We consider previous $r$ votes to refer to those that are selected from the voters' network.
Therefore, at time $t$, $r$ previous votes are 
 the number of votes for $C_0$ and $C_1$, represented by $c_0^r
(t)$ and $c^r_1(t)$.
Hence, $c_0^r(t)+c_1^r(t)=r$ holds.
If $r > t$, voters can view $t$ previous votes for each candidate.
 In the limit $r \rightarrow \infty$, voters can view all previous votes \cite{Hisakado3}.
We define the number of  previous votes for $C_0$ and $C_1$ as
 $c_0^\infty (t)\equiv c_0(t)$ and $c_1^\infty(t)\equiv c_1(t)$.
In the real world, the number of references $r$ depends on  the voters; however, we considered $r$  constant in this study.

The herder is considered  a digital herder \cite{Hisakado3}.
We define $c(t)^r_1/r=1-c(t)^r_0/r=Z_r (t)$.
A digital herder's behavior is defined by the function $f(z)=\theta(Z_r-1/2)$, where
$\theta(Z)$ is a Heaviside function,
 and the threshold was 1/2, corresponding to the majority decision.
In \cite{W2}, a model in which the threshold was different from 1/2 was discussed.
The model corresponds to voting without independent voters.
We define $P_{h1}^{r}(t)$ as the probability that the   herder  would vote for $C_1$ at time $t$,
\begin{equation}
P_{h1}^{r}(t)=\left\{ 
\begin{array}{ll}
1&;c_1^r(t)>r/2, \nonumber \\
1/2&;c_1^r(t)=r/2, \nonumber \\
0 &:c_1^r(t)<r/2 \nonumber.
\end{array} 
\right \}
\label{herd}
\end{equation} 

Independents and herders appear randomly and vote.
We set the ratio of independents  to 
herders as $(1-p)/p$.
In this study, we focus mainly on the upper limit of $t$, which
refers to voting by an infinite number of voters.

\section{IV. Voting model on Network}

\begin{figure}[h]
\includegraphics[width=110mm]{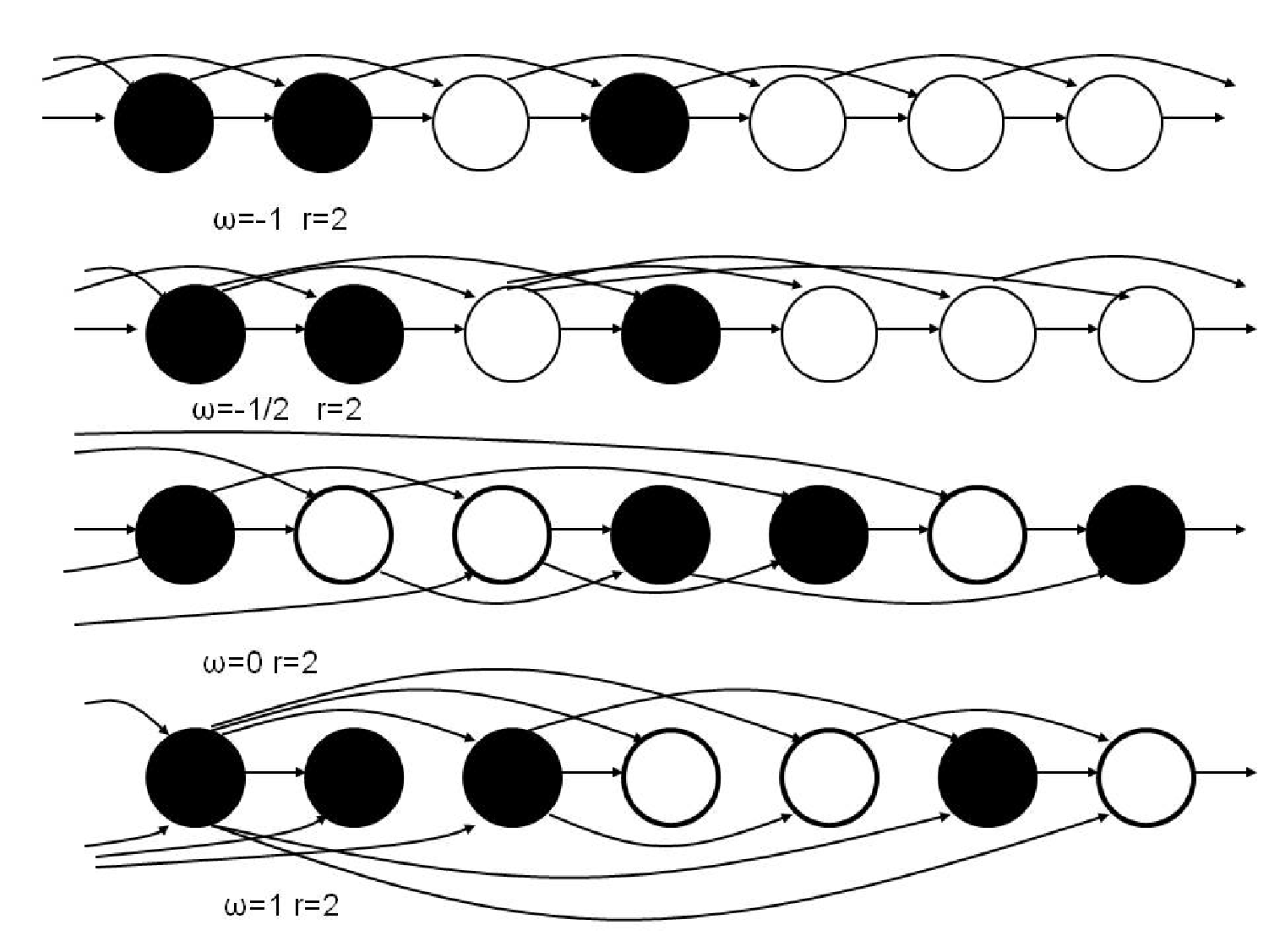}
\caption{Representation of graphs.
An extended 1D lattice, Fermion-like graph \cite{Hisakado7}, random graph, and BA model, for $\omega=-1, -1/2,0,1$, $r=2$, respectively.
The white (black) dot is a voter who voted for candidate $C_0 (C_1)$. 
The two arrows pointing toward a dot represent a voter who refers to two voters when $r=2$.
In the case of an extended 1D lattice, a voter refers to the latest two voters.
In the case of a random graph, a voter refers to two previous voters who are selected randomly.
In the case of the BA model, a voter refers to two previous voters who are selected via popularity.
Hence, there are voters who play the role of a hub in the BA model.
The first voter in the network graph is a hub that influences many other voters.}
\label{graph}
\end{figure}

In this section, we combine the network and voting models discussed in the previous two sections.  
We consider the voter  to be able to view $r$ previous votes.
In Section II, we define how to select previous voters for reference.
The influence of reference voters is represented as a voting model in a network.
Our problem lies in how the network affects the voting model. 
In this study, we analyzed the cases by comparing models based on
networks created by a parameter $\omega$.
The network includes a random graph, the  BA model case, and the extended lattice, which correspond to $\omega=1,0,-1$, respectively.
In Fig.\ref{graph}, we illustrate each  of these three cases for $r=2$.
A white (black) dot indicates a voter who voted for candidate $C_0 (C_1)$. 
The two arrows point toward a dot, indicating that a voter refers to two other voters when $r=2$.
In the case of a 1D extended lattice, the voter refers to the latest two voters.
In the case of a random graph, a voter refers to two previous voters who are  selected randomly.
In the case of the BA model, a voter refers to two previous voters selected through the voter's popularity network.
A positive $\omega$ network has the characteristics of a scale-free network with hubs.
However, in the  negative $\omega$ network, the variance in the number of connections in each node  is small.

We define $P_1^{r}(t)$ as the probability that the $t$-th voter would vote for $C_1$ using Eq.(\ref{herd}), 
\begin{equation}
P_{1}^{r}(t)=\left\{ 
\begin{array}{ll}
p+(1-p)q&;c_1^r(t)>r/2, \nonumber \\
p/2+(1-p)q&;c_1^r(t)=r/2, \nonumber \\
(1-p)q &:c_1^r(t)<r/2 \nonumber.
\end{array} 
\right \}
\label {ge}
\end{equation} 
It is the sum of the voting of herders and independents.

In the scaling limit, $t=c_0(t)+c_1(t)=c_0^{\infty}+c_1^{\infty}\rightarrow \infty$,
 we define
\begin{equation}
Z(t)=\frac{c_1(t)}{t}
\Longrightarrow Z_{\infty}.
\label{Z}
\end{equation}
$Z(t)$ is the ratio of voters who vote for $C_1$ at $t$.

Here, we define $\pi(Z)$ as the majority probability of binomial distributions of $Z$,
in other words, the probability that $c_1^r(t)>1/2$.
When $r$ is odd,
\begin{equation}
\pi(Z)=\sum_{g=\frac{r+1}{2}}^{r}
\left(
\begin{array}{cc}
r\\
g 
\end{array}
\right)
Z^g(1-Z)^{r-g}.
\label{2}
\end{equation}

$\pi$ can be calculated as follows:
\begin{equation}
\pi(Z)=\frac{(2n+1)!}{(n!)^2}\int_0^{Z}x^n(1-x)^ndx=\frac{1}{B(n+1,n+1)}\int_0^{Z}x^n(1-x)^ndx.
\label{pi}
\end{equation}
Eq.(\ref{pi}) can be applied when the referred voters overlap.
In fact, the referred voters do not select an overlap.
However, in this study, we used this approximation to investigate the large $t$ limit.

Each popularity has a color in the voting model.
The color depends on whether the voter voted for candidate $C_1$ or $C_0$.
In Fig.\ref{BA}, voters who voted for $C_1$($C_0$) are represented by black (white) circles.

We define the total popularity of voters who vote for candidate $C_1$($C_0$) at $t$ as $g_1(t)$($g_0(t)$).
In the scaling limit, $g_0(t)+g_1(t)=(1+\omega)r(t-r+1)\rightarrow \infty$,
we define
\begin{equation}
\frac{g_1(t)}{(1+\omega)r(t-r+1)}=\hat{Z}(t)\Longrightarrow \hat{Z}_{\infty}.
\end{equation}
$\hat{Z}(t)$ is the ratio of popularity  for $C_1$ at $t$.

We can denote the evolution of popularity as
\begin{eqnarray}
& &g_1(t)=\hat{k} \rightarrow \hat{k}+i:
\nonumber \\
& r/2\leq j\leq r, i=\omega j+r,& P_{\hat{k},t}(i)={}_{r}C_{j}\hat{Z}^{j}(1-\hat{Z})^{r-j}[(1-p)q+p],
\nonumber \\
& 0\leq j< r/2, i=\omega j+r, & P_{\hat{k},t}(i)={}_{r}C_{j}\hat{Z}^{j}(1-\hat{Z})^{r-j}(1-p)q,
\nonumber \\
& r/2\leq j\leq r, i=\omega j, & P_{\hat{k},t}(i)={}_{r}C_{j}\hat{Z}^{j}(1-\hat{Z})^{r-j}(1-p)(1-q),
\nonumber \\
& 0\leq j< r/2, i=\omega j, & P_{\hat{k},t}(i)={}_{r}C_{j}\hat{Z}^{j}(1-\hat{Z})^{r-j}[(1-p)(1-q)+p],
\nonumber \\
\label{pd}
\end{eqnarray}
where $P_{\hat{k},t}(i)$  is the probability of the transition  from $\hat{k}$ to
$\hat{k}+i$ at $t$.
\subsection{A. Information cascade transition}
Here, we consider self-consistent equations for popularity at a large $t$ limit,
\begin{equation}
(1+\omega) r\hat{Z}_{\infty}=\sum_{j=1}^{r}P_{\hat{k},t}(i)\cdot i=r(1-p)q+rp\pi(\hat{Z}_{\infty})+r\omega\hat{Z}_{\infty}.
\end{equation}
Hence, we can obtain
\begin{equation} 
\hat{Z}_{\infty}=(1-p)q+p\pi(\hat{Z}_{\infty}).
\label{sc1}
\end{equation}
This is a self-consistent equation that does not depend on $\omega$.
The self-consistent equation for voting ratio $Z_{\infty}$ is also
\begin{equation} 
Z_{\infty}=(1-p)q+p\pi(Z_{\infty}).
\label{sc2}
\end{equation}

We define a new variable $\hat{\Delta}_t$ such that
\begin{equation}
\hat{\Delta}_t=g_1(t)-r(t-r+1)=\frac{1}{2}\{g_1(t)-g_0(t)\}.
\label{d}
\end{equation}
For convenience, we change the notation from $\hat{k}$ to $\hat{\Delta}_t$.
Therefore, $\hat{\Delta}_t$ holds within  $\{-r(t-r+1),r(t-r+1)\}$. 
Given $\hat{\Delta}_t=\hat{u}$, we obtain a random walk model:
\begin{eqnarray}
& &\hat{\Delta}=\hat{u} \rightarrow \hat{u}+i:
\nonumber \\
& r/2\leq j\leq r, i=(r+(2j-r)\omega/(1+\omega),& P_{\hat{u},t}(i)={}_{r}C_{j}\hat{Z}^{j}(1-\hat{Z})^{r-j}[(1-p)q+p],
\nonumber \\
& 0\leq j<r/2, i=(r+(2j-r)\omega/(1+\omega),& P_{\hat{u},t}(i)={}_{r}C_{j}\hat{Z}^{j}(1-\hat{Z})^{r-j}(1-p)q,
\nonumber \\
& r/2\leq j\leq r, i=(r+(2j-r)\omega/(1+\omega),& P_{\hat{u},t}(i)={}_{r}C_{j}\hat{Z}^{j}(1-\hat{Z})^{r-j}(1-p)(1-q),
\nonumber \\
& 0\leq j< r/2, i=(r+(2j-r)\omega/(1+\omega),& P_{\hat{u},t}(i)
={}_{r}C_{j}\hat{Z}^{j}(1-\hat{Z})^{r-j}[(1-p)(1-q)+p],
\nonumber \\
\label{pd2}
\end{eqnarray}
where $\hat{Z}=\hat{k}/(2r(t-r+1))=\hat{u}/(2r(t-r+1))+1/2$ 
and
$P_{\hat{u},t}(i)$ is the probability of the transition  from $\hat{u}$ to $\hat{u}+i$ at $t$.


We now consider the continuous limit $\epsilon \rightarrow 0$
\begin{equation}
\hat{X}_\tau=\epsilon\hat{\Delta}_{[\tau/\epsilon]},
\end{equation}
where $\tau=t\epsilon$. 
Approaching the continuous limit, we obtain the following stochastic partial differential equation: 
\begin{eqnarray}
\textrm{d}\hat{X}_\tau&=&[\frac{2r(1-p)q}{1+\omega}-\frac{r}{1+\omega}+\frac{\omega}{(1+\omega)}\frac{\hat{X}_{\tau}}{(\tau-r+1)}
\nonumber \\
& &
+\frac{2rp}{1+\omega}\frac{(2n+1)!}{(n!)^2}\int_0^{\frac{1}{2}
+\frac{\hat{X}_\tau}{2r(\tau-r+1)}}x^n(1-x)^ndx ]\textrm{d}\tau+\sqrt{\epsilon}{\rm d}B,
\label{ito0n}
\end{eqnarray}
where ${\rm d}B$ is the Wiener process.
For $r=1$, the equation becomes:
\begin{equation}
\textrm{d}\hat{X}_\tau=[\frac{(1-p)(2q-1)}{1+\omega}+\frac{p+\omega}{1+\omega}\frac{\hat{X}_\tau}{\tau}]\textrm{d}\tau+\sqrt{\epsilon}{\rm d}B.
\label{ito01}
\end{equation}
The derivation of Eq.(\ref{ito0n})    is the extension of  the previous studies and see  \cite{Hisakado5} in detail.

The relation between the voting ratio for $C_1$ and $\hat{X}_\infty$ is 
\begin{equation}
\frac{\hat{X}_\infty}{2r(\tau-r+1)}=\hat{Z}_{\infty}-
\frac{1}{2}.
\label{xz}
\end{equation}

We can assume that the stationary solution is
\begin{equation}
\hat{X}_\infty=r\bar{\hat{v}}\tau+r(1-p)(2q-1)\tau,
\label{hnn}
\end{equation}
where $\bar{\hat{v}}$ denotes a constant.
As Eq.(\ref{xz}) and $0\leq\hat{Z}\leq1$, we obtain
\begin{equation}
-1\leq\bar{\hat{v}}+(1-p)(2q-1)\leq1.
\label{hani}
\end{equation}
Substituting Eq.(\ref{hnn}) into Eq.(\ref{ito0n}), we obtain
\begin{equation}
\bar{\hat{v}}=-p+\frac{2p\cdot(2n+1)!}{(n!)^2}\int_0^{\frac{1}{2}+\frac{(1-p)(2q-1)}{2}+
\frac{\bar{\hat{v}}}{2}} x^n(1-x)^ndx.
\label{i2}
\end{equation}
This equation is a self-consistent equation.
Eq.(\ref{i2}) does not depend on parameter $\omega$.
Hence, the transition point  is the same for all $\omega> -1$.

When $p\leq p_c$, the phase is referred to as the peak phase and the solution is $\bar{\hat{v}}_0$.
Eq.(\ref{i2}) admits three solutions for $p>p_c$:
When $p>p_c$, the upper and lower solutions  
are stable; on the contrary, the intermediate solution is unstable.
The two stable solutions correspond to good $\bar{\hat{v}}_+$ and bad equilibria $\bar{\hat{v}}_-$, respectively, 
and the distribution becomes the sum of  the two Dirac measures.
This is a two-peak phase.

The phase transition point, $p_c$ is common for all models.
If $r=2n+1\geq3$, then a phase transition occurs in the range $0\leq p\leq1$.
If a voter obtains information from either one or two voters, there is no phase transition.

\subsection{B. super-normal transition}
We expand $\hat{X}_\tau$ around solution $\bar{\nu}=r\bar{\hat{v}}\tau+r(1-p)(2q-1)\tau$,
\begin{equation}
\hat{X}_\tau=r\bar{\hat{v}}\tau+r(1-p)(2q-1)\tau+r\hat{W}_\tau.
\label{w2}
\end{equation}
We set $\hat{X}_\tau\gg \hat{W}_\tau$.
This result indicates that $\tau\gg 1$.
We rewrite Eq.(\ref{ito0n}), using Eq.(\ref{w2}) and obtain the following:
\begin{eqnarray}
\textrm{d}\hat{W}_\tau
&=&[\frac{\omega+pA}{1+\omega}]\frac{\hat{W}_\tau}{\tau}\textrm{d}\tau+\sqrt{\epsilon}{\rm d}B,
\label{ito4}
\end{eqnarray}
where 
\begin{equation}
A=\frac{(2n+1)!}{(n!)^2\cdot 2^{2n}}(1-\{\bar{\hat{v}}+(1-p)(2q-1)\}^2)^n.
\end{equation}

The trend in this solution is $(\omega+pA)/(1+\omega)$ where: 
there is a transition in the trend at the 
$(\omega+pA)/(1+\omega)=1/2$. 
We set the solution to this equation as $p_{vc}$.
When $(\omega+pA)/(1+\omega)>1/2$, the convergence speed is slower than that in the normal case.
However, when $(\omega+pA)/(1+\omega)<1/2$, the convergence speed is slower than that  in the normal case.
We call this transition a super-normal transition, and the transition point is $p_{vc}$.

When $\omega\geq 1$, $(\omega+pA)/(1+\omega)\geq1/2$ because $A\geq 0$.
Thus, there is no super-normal transition at $\omega\geq 1$.
We observed a super-normal transition at 
$1>\omega> -1 $.

In summary, the convergence of  $Z(t)$ is 
\begin{equation}
V(Z(t))\propto
\left\{
\begin{array}{ccc}
t^{-1} & p<p_{vc}, \\
t^{(2pA-2)/(1+\omega)} & p>p_{vc}, \\
\frac{\log(t)}{t} & p=p_{vc},
\end{array}
\right.
\label{sn}
\end{equation}
where $V(Z(t)$ is the variance of $Z(t)$.
The analysis of  super-normal transition  in detail  is  in Appendix F.

\subsection{C. Symmetric case, $q=1/2$}
Here, we consider a symmetric model $q=1/2$.
When $r\geq3$, there are two stable solutions and one unstable solution $\bar{\hat{v}}=0$ above $p_c$.
The  vote ratio for $C_1$ is
a good or  bad equilibrium.
In one sequence, $Z$ 
is taken as $\bar{\hat{v}}/2+1/2$ in the case of good equilibrium or
as $-\bar{\hat{v}}/2+1/2$ in the case of  bad equilibrium 
where $\bar{\hat{v}}$ is the solution of Eq.(\ref{i2}).
This indicates a two-peak phase, and 
 the critical point is $p_c=\frac{(n!)^2}{(2n+1)!}2^{2n}=1/A$, where the gradient of the RHS of Eq.(\ref{i2}) at $\bar{\hat{v}}=0$ is $1$.
In the case of $r=3 (n=1)$, $p_c=2/3$ and $r=5(n=2)$, $p_c=8/15$.
As $r$ increases, $p_c$ moves toward $0$.
In the large limit, $r\rightarrow \infty$, $p_c$ becomes $0$.
This is consistent with the case where herders obtain information from all previous voters \cite{Hisakado3}.
The discussion above does not depend on $\omega$.
In the limit $\omega=-1$, there is no phase transition in the lattice case.

We consider the super-normal transition in the symmetric case $q=1/2$ by considering the case $r=2n+1\geq3$.
In this case, we observe an information cascade transition.
If $r\leq 2$, no information cascade transition is  observed.

In one-peak phase $p\leq p_c$, the only solution is $\bar{\hat{v}}=0$.
$p_c$ is the critical point of   the information cascade transition. 
The critical point of  convergence is $p_{vc}=(1-\omega)/2A=\frac{(1-\omega)}{2}p_c$.

Above $p_c$, in the two-peak phase, we obtain two stable solutions that are not $\bar{\hat{v}}=0$.
At $p_c$, $\bar{\hat{v}}$ moves from $0$ to one of  the two stable solutions.
In one voting sequence,  the votes converge to one of these stable solutions.
Here, $\bar{\hat{v}}$ is the solution of Eq.(\ref{i2}).
In the case $r=3$ we obtain $\bar{\hat{v}}=\pm\sqrt{(3p-2)/p}$ and
$p_{c}=2/3$. 
The critical point of convergence in the two-peak phase was $p_{vc}=(5+\omega)/6$.

\begin{figure}[h]
\includegraphics[width=110mm]{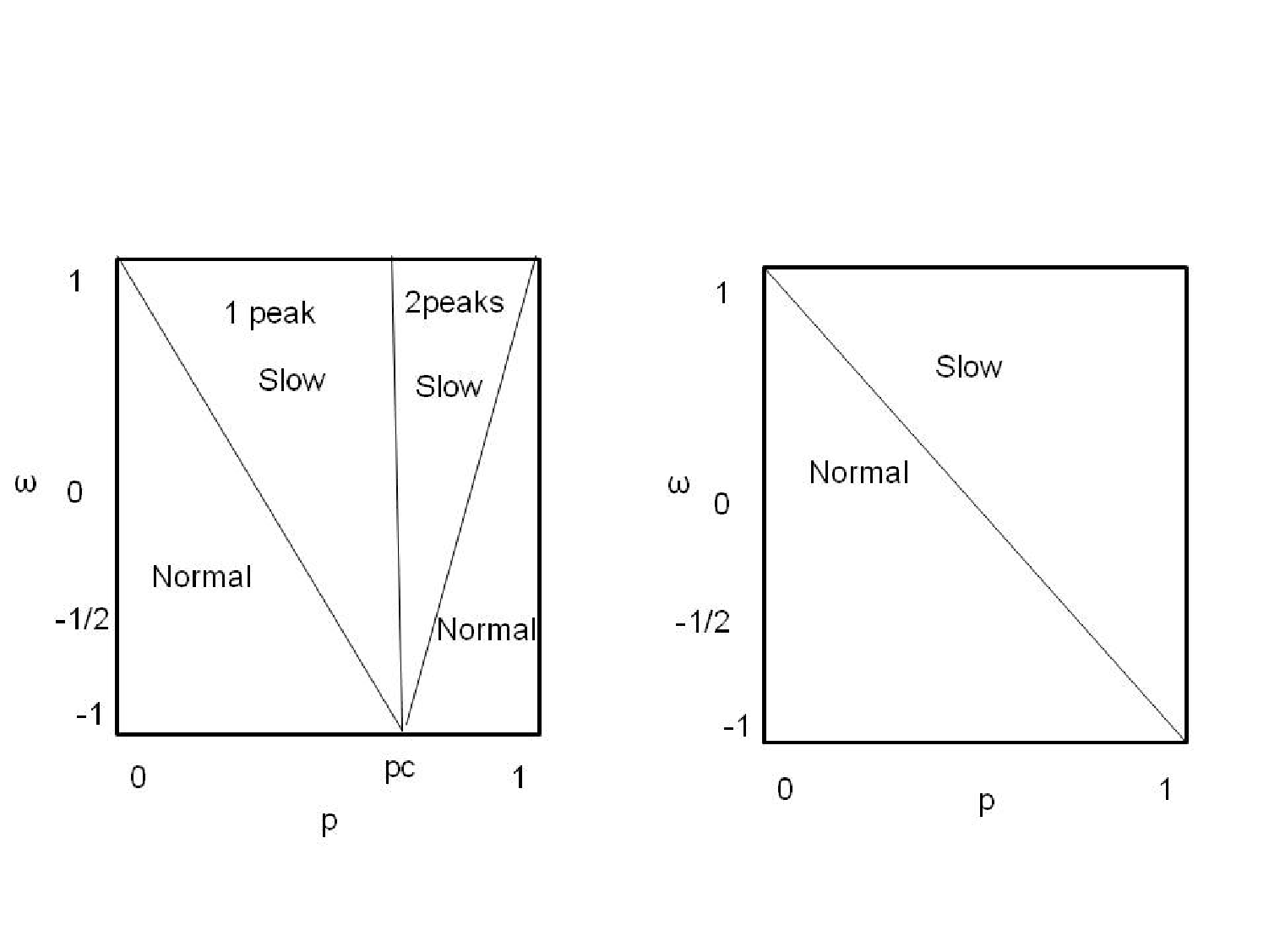}
\caption{Phase diagram of the symmetric case ($q= 1/2)$ for $r=1$ and $r=3$ cases. In the case of $r=3$, $p_c$ is 2/3.
The horizontal axis is $p$, and the vertical axis is $\omega$.}
\label{phase}
\end{figure}

In Fig.\ref{phase} 
we show the phase diagrams for the cases $q=1/2$, $r=3$, and $r=1$.
In the case $r=1$, there is only a super-normal transition.
The super-normal transition disappears  at $\omega=-1$.
In the case $r=3$,  we can confirm two types of phase transitions, the information cascade transition and
a super-normal transition.
The super-normal transition disappears at $\omega=-1$.
In the analog herder case, the phase diagram becomes the same as that  in the $r=1$ case.
Analog herders vote for each candidate with probabilities proportional to the candidates' votes.

\subsection{D. Asymmetric case, $q\neq 1/2$ }

\begin{figure}[h]
\includegraphics[width=110mm]{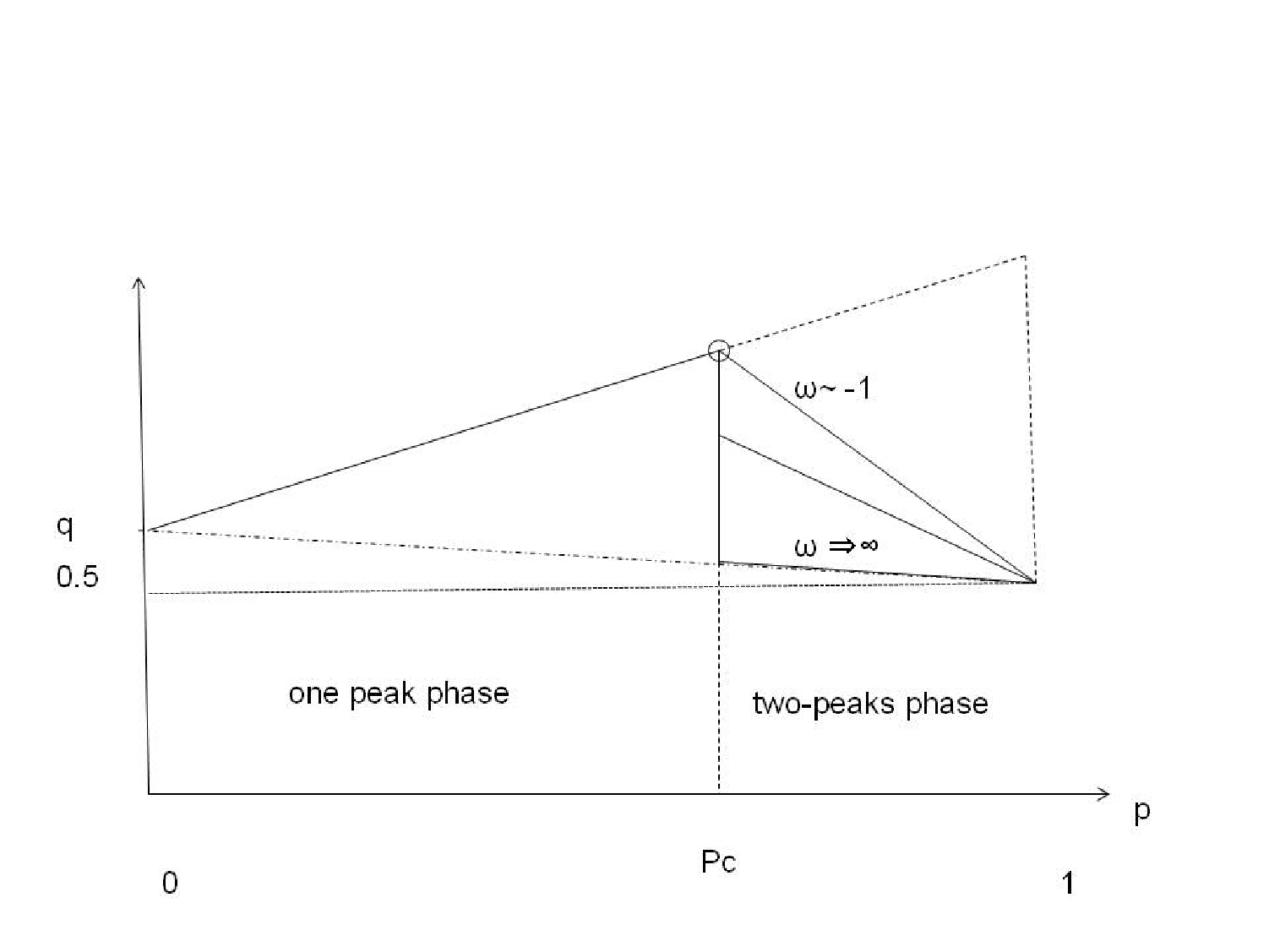}
\caption{Image of expected correct ratio and first order phase transition for the asymmetric case. The effects of the hubs are observed in the correct ratio in the two-peak phase.
As the effects of the hub increases, the gap also increases. The numerical simulation of expected correct ratio is in Fig.\ref{fig:gamma_q06_and_EZ}.} 
\label{order}
\end{figure}

Next, we consider the asymmetric case.
In this case, the phase transition is of the first order that
exhibits a discontinuity at the critical point $p_c$.
We investigate the correct ratio for this phase transition.
To define the correct ratio, we assume that candidate 1(0) is the correct(wrong) one.
The correct ratio of independent voters is $q>1/2$, and
the correct ratio of the herder determines the total correct ratio.
Using the correct ratio, we can confirm the voting performance.
The phase transition has a negative effect on the performance.
In Fig.\ref{order}, we show the image of the correct ratio in several $\omega$.
The horizontal axis represents $p$ and the vertical axis represents the correct ratio.
We observe a gap in the correct ratio at 
the transition point $p_c$,  when $\omega>-1$
 $p_c$ does not depend on $\omega$.
When $\omega=-1$, no phase transition occurred.

In the limit $\omega\rightarrow \infty$,
the trend of Eq.(\ref{ito4}) is 1 for all $Z$.
In this case, the hubs have the maximum number of links.
At the critical point,  $p_c$ the correct ratio depends only on the initial voting.
The two equilibria are $\bar{\hat{v}}_+$ and $\bar{\hat{v}}_-$.
The probabilities of good and bad equilibria are
$\alpha $ and $\beta$.
The correct ratio is $Z_{\infty}=(\alpha\bar{\hat{v}}_+ + \beta\bar{\hat{v}}_- )/2+1$.  
In the case $\omega\rightarrow \infty$, we obtain $Z_{\infty}=(\alpha\bar{\hat{v}}_+ +\beta\bar{\hat{v}}_- )/2+1= p/2+(1-p)q$, which is the dotted bottom line in Fig.\ref{order}.
In general, for $\omega>-1$, we obtain 
\begin{equation}
 \lim_{p\rightarrow p_c}\bar{v}= 2\bar{\hat{v}} -1 >(\alpha\bar{\hat{v}}_+ + \beta\bar{\hat{v}}_- )/2+1 \geq p/2+(1-p)q. 
\end{equation}

In summary, in the one-peak phase, the correct ratio does not depend on $\omega$.
In contrast, in the two-peak phase, the correct ratio depends on $\omega$. 
As $\omega $ increased, the gap at the transition
point increases and the correct ratio decreases.
In other words, the correct ratio of the herders decreases.
In the lattice case $\omega=-1$, there is no phase transition, and the correct ratio increases without the phase transition as $p$ increases.
The lattice case is summarized in Appendix C and  discussed in \cite{Hisakado4}.
The effects of  hubs are observed in the correct ratio during the two-peak phase.
As the hub effects increase, the gap  also increases.

\section{V. Numerical Study of Phase Transitions}

In this section, we present the results of  numerical studies on
phase transitions of the voting model in  networks.
We adopt the model parameters $(r,q)$ as $r=3$ and $q\in \{0.5,0.6\}$
and change the control parameter $p\in [0,1]$.
With regard to $\omega$, we adopt $\omega\in \{-0.9,-2/3,-1/2,0,1,2,9\}$.
We estimate the statistical quantities of the model using  simple
Monte Carlo sampling. Using the probabilistic rule of the network and 
voting model, we generated $10^2$ networks and 
$10^4$ sample sequences in each network.
After estimating the statistical quantities by taking the average over
the sample sequences, we calculated the average of the networks.
To estimate the universal function of a continuous transition,
we generated $10^2$ and $10^4$ sample sequences.

When $\omega=-1$, we adopt an extended lattice with $r=3$, which
is different from that of the network with $\omega=-1$ because of the initial condition of the network.
We used a perfect network as the initial condition for this numerical simulation.


 
Probability that voter 1  was chosen by the next voters is zero because the  in-degrees  
is 0. In addition, the in-degrees and popularity  of Voters 2 and 3
are 1 and 2, respectively. Because they are smaller than those of the
voters for $t>3$, they are chosen less by the next voters.

\subsection{A. Symmetric case, $q=1/2$}
In this subsection, we consider the symmetric case, $q=1/2$.
A phase transition occurred at $p=p_{c}=2/3$.
In the one-peak phase $p<p_{c}=2/3$, $Z(t)$
converges to the unique solution to Eq.(\ref{sc2}).
There is also a super-normal transition at
$p=p_{vc}=(1-\omega)/3$.
As Eq.(\ref{sn}), the power-law exponent $\gamma$ of $V(Z(t))\propto t^{-\gamma}$ is given by 
\[
\gamma=
\left\{
\begin{array}{cc}
1 & p<p_{vc}, \\
(2-3p)/(1+\omega) & p>p_{vc}, 
\end{array}
\right.
\]
where $V(Z(t))$ is the variance of $Z(t)$.
At $p= p_{vc}$, $V((Z(t))\propto \log t/t$.
Here, $\gamma=1$ is the normal phase and $0<\gamma<1$ is the super diffusion phase.
We estimate $\gamma$ using the following estimator:
$\hat{\gamma}=\log_{2} V(Z(T/2))/V(Z(T))$.

\begin{figure}[htbp]
\begin{center}
\begin{tabular}{cc}
\includegraphics[width=8cm]{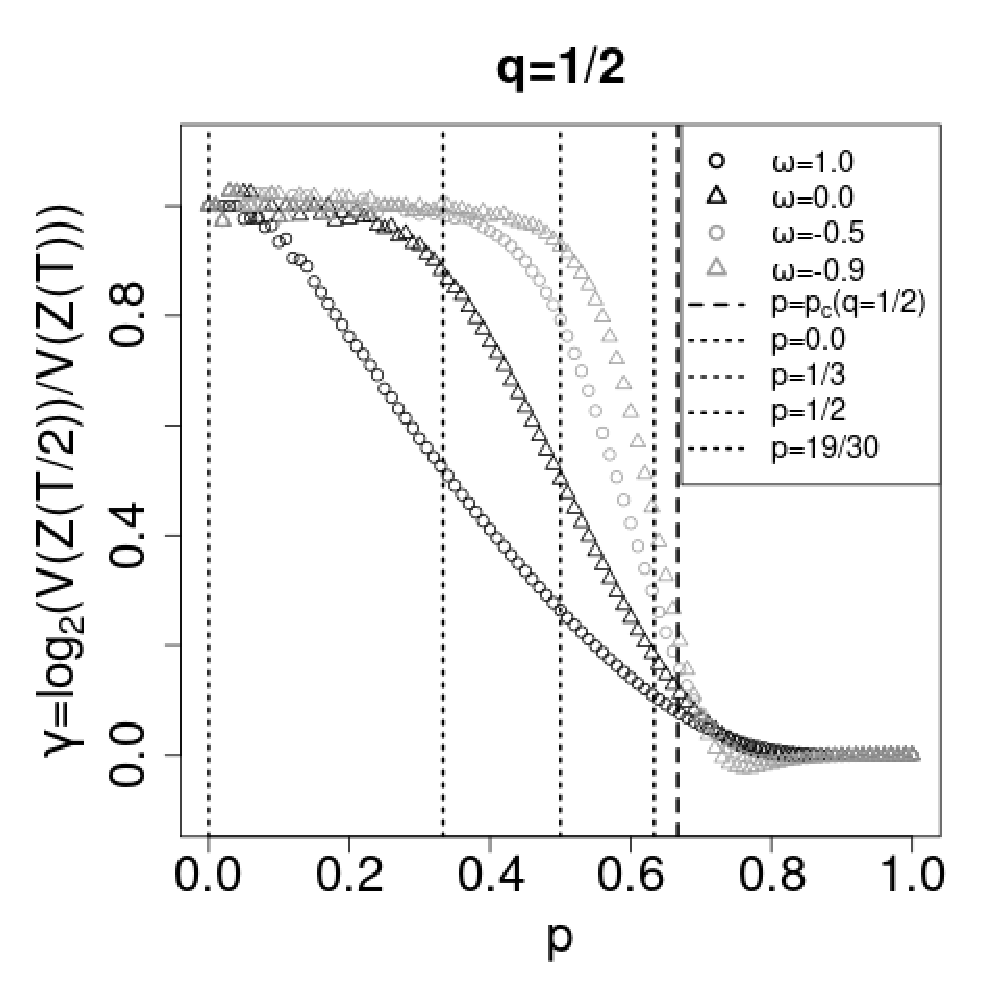} & 
\includegraphics[width=8cm]{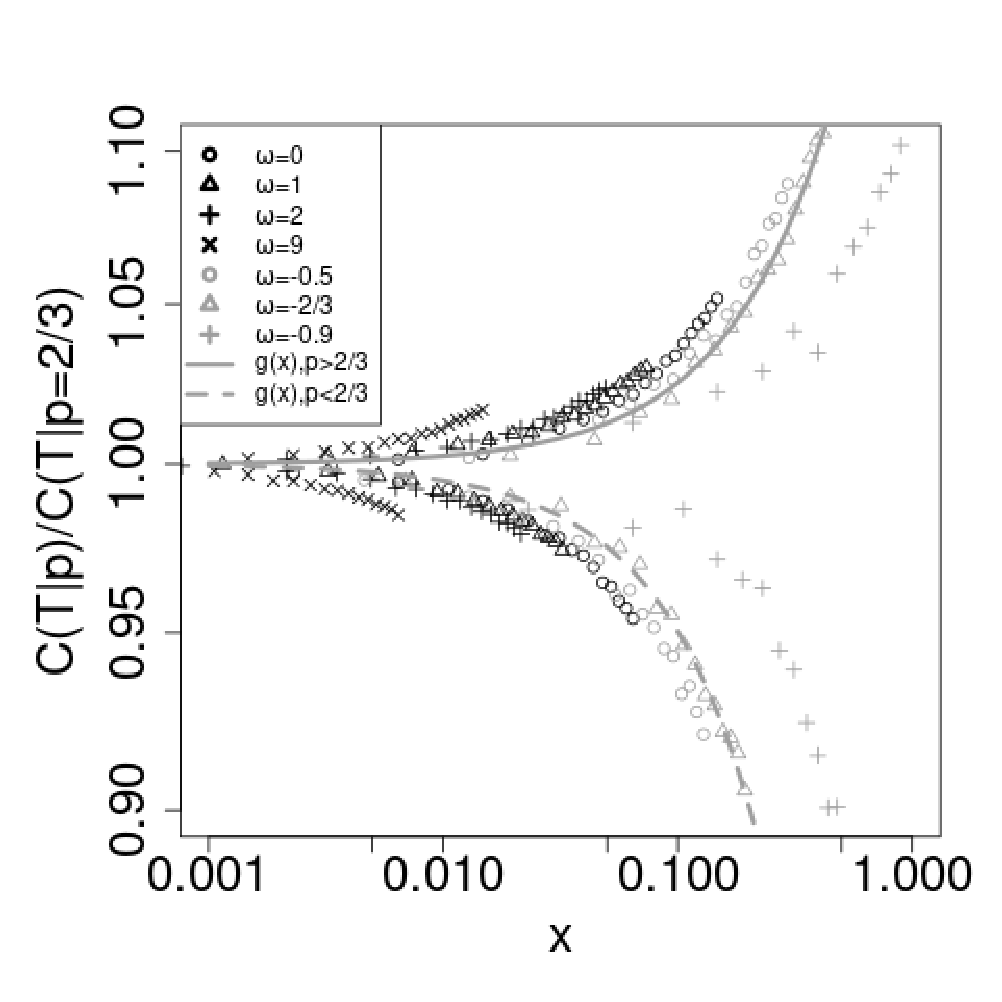}
\end{tabular}
\end{center}
\caption{Left:
Plot $\gamma$ vs. $p$ for $q=0.5, r=3$ and $T=10^{4}$,
$\omega=1(\mbox{black} \bigcirc),0.0(\mbox{black} \bigtriangleup),-0.5(\mbox{gray} \circ)$
and $-0.9(\mbox{gray} \bigtriangleup)$,
where $T$ is the time.
The broken and dotted lines show the value of $p_c =2/3$ and
$p_{vc}(q,\omega)$ for $\omega\in \{1,0.0,-0.5,-0.9\}$.
Right: Plot of the universal function and the estimated ones.
For $\omega\in \{0(\mbox{black} \bigcirc),1(\mbox{black} \bigtriangleup),2(\mbox{black} +),9(\mbox{black} \times),-0.5(\mbox{gray} \bigcirc),-2/3(\mbox{gray} \bigtriangleup),-0.9(\mbox{gray} +)\}$ we plot $C(T|p)/C(T|p_{c})$
as function of $x$ in Eq.(\ref{eq:x}).
The gray lines show the universal function
$g(x)$ of Eq.(\ref{eq:x}) for $p>p_{c}$(gray solid) and $p<p_{c}$(gray broken).
}
\label{fig:gamma_q05_and_UC}
\end{figure}	

The left figure of Fig.\ref{fig:gamma_q05_and_UC}
plots $\gamma$ vs. $p$ for $\omega\in \{1,0,-0.5,-0.9\}$.
The horizontal axis represents the ratio of herders $p$.
The vertical axis represents the speed of convergence $\gamma$.
The vertical broken line indicates the critical point, $p_{c}(1/2)=2/3$.
The other vertical dotted lines show $p_{vc}$ for $\omega\in \{1,0,-0.5,-0.9\}$,
respectively. 
The phase transition
occurs at $p=p_{c}(1/2)$.
For $p>p_{c}$, $\gamma$ is almost zero and
the variance in $Z(t)$ becomes constant. 
There are two stable states.
For $p>p_{c}$, $V(Z(t))$ does not decrease to zero.
For $p<p_{c}$, a super-normal transition occurs. 
The positions of the
critical points are vague; evidently, the region $p$ for the
normal phase for $\omega=1$ is extremely narrow, and it considers $p_{vc}(\omega=1)=0$. 
For $\omega=0$, $p_{vc}=1/3$ and a plateau region was observed for $p<0.25$.
As $\omega$ decreases from $\omega=1\to -0.9$, the plateau region widens.
Near the critical point $p_{vc}$, $\gamma$ is smaller than one, and estimating $p_{vc}$ from the figure is difficult.

To study the universality class of continuous phase transitions
at $p_{c}$, we examined the order parameter $C(T)$. $C(T)$ is defined as follows,
\[
C(t|p)=E(c_1 (t+1)|c_1 (1)=1)-E(c_1 (t+1)|c_1 (1)=0),
\]
and reflects the sensitivity of the initial condition for the stochastic process.

\begin{figure}[h]
\includegraphics[width=110mm]{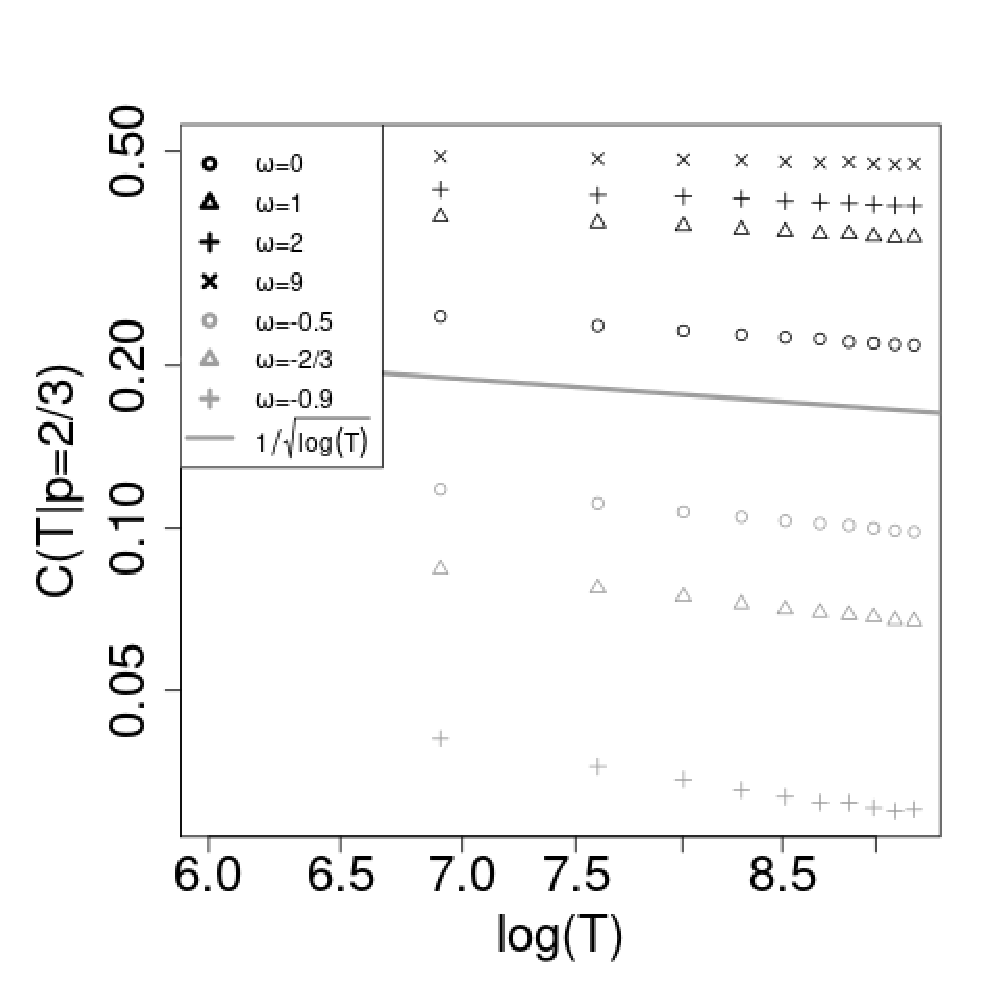}
\caption{Plot $C(T)$ vs Log $T$ at $\omega=9,2,1,0, -0.5,-2/3,-0.9$. $\alpha=1/2$ and does not depend on $\omega$. }
\label{alpha}
\end{figure}

In Fig.\ref{alpha}, we show the critical exponent $\alpha=1/2$ and it does not depend on  $\omega$.

As discussed in Appendix D, the universal function $g(x)$
of the continuous phase transition for the symmetric case is defined as
\begin{equation}
g(x)=\lim_{t\to \infty}C(t|p)/C(t|p_{c})
=\left\{
\begin{array}{cc}
\sqrt{\frac{x}{1-e^{-x}}} & p\to p_{c}-0,t\to \infty, x=\frac{3(p_c-p)/2}{1+\omega}\ln t, \\
\sqrt{\frac{2x}{e^{2x}-1}} & p\to p_{c}+0,t\to \infty,x=\frac{3(p-p_c)}{1+\omega}\ln t. \\ 
\end{array}
\right. \label{eq:x}
\end{equation}

\subsection{B. Asymmetric case, $q=0.6$}

A phase transition occurred at $p=p_{c}(q=0.6)=0.781$. 
As in the symmetric case $Z(t)$
converges to the unique solution to 
Eq.(\ref{sc2}) for one-peak
phases for $p<p_{c}$.
The variances of $Z(t)$ and $V(Z(t))$ also decrease to zero in the limit.
There is also a super-normal transition at
$p=p_{vc}(0.6)$. When $\omega=1(0)$, $p_{vc}=0(0.360)$.
When $\omega=-0.5,-0.9$, only the normal convergence phase exists.

\begin{figure}[htbp]
\begin{center}
\begin{tabular}{cc}
\includegraphics[width=8cm]{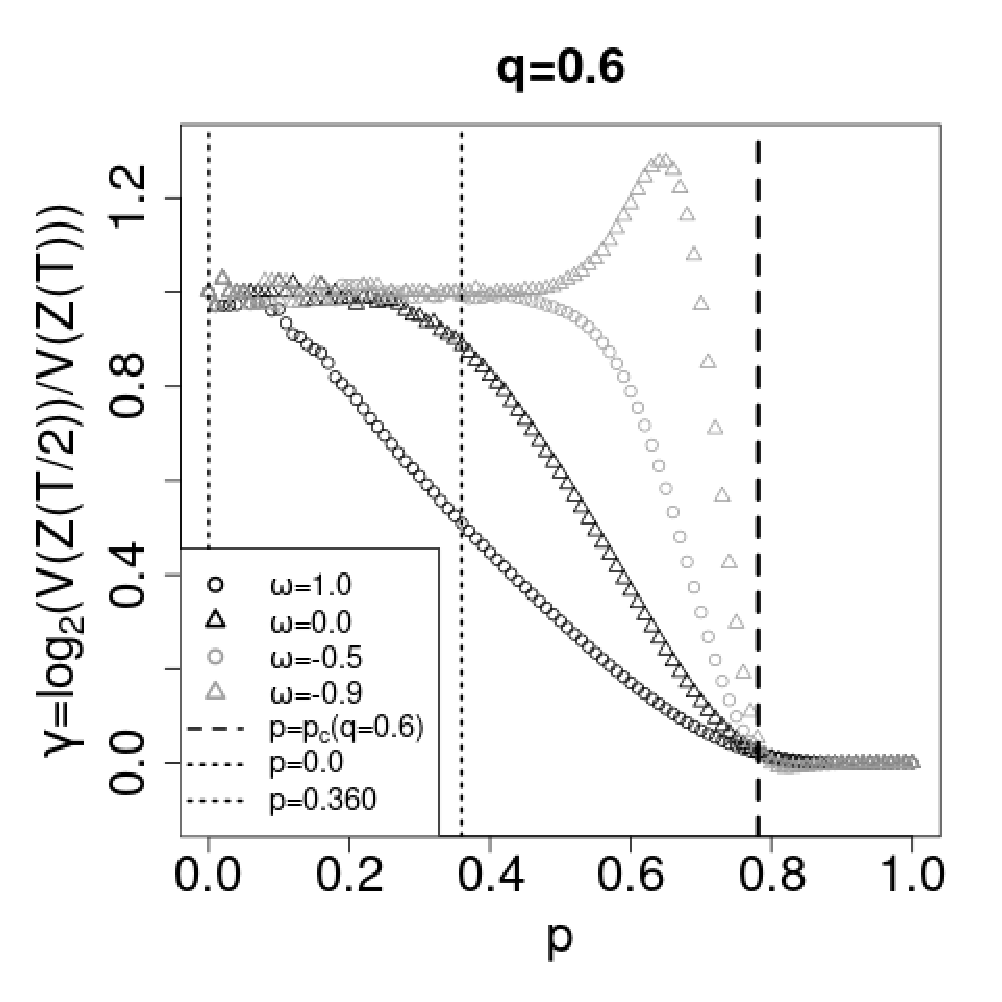} &
\includegraphics[width=8cm]{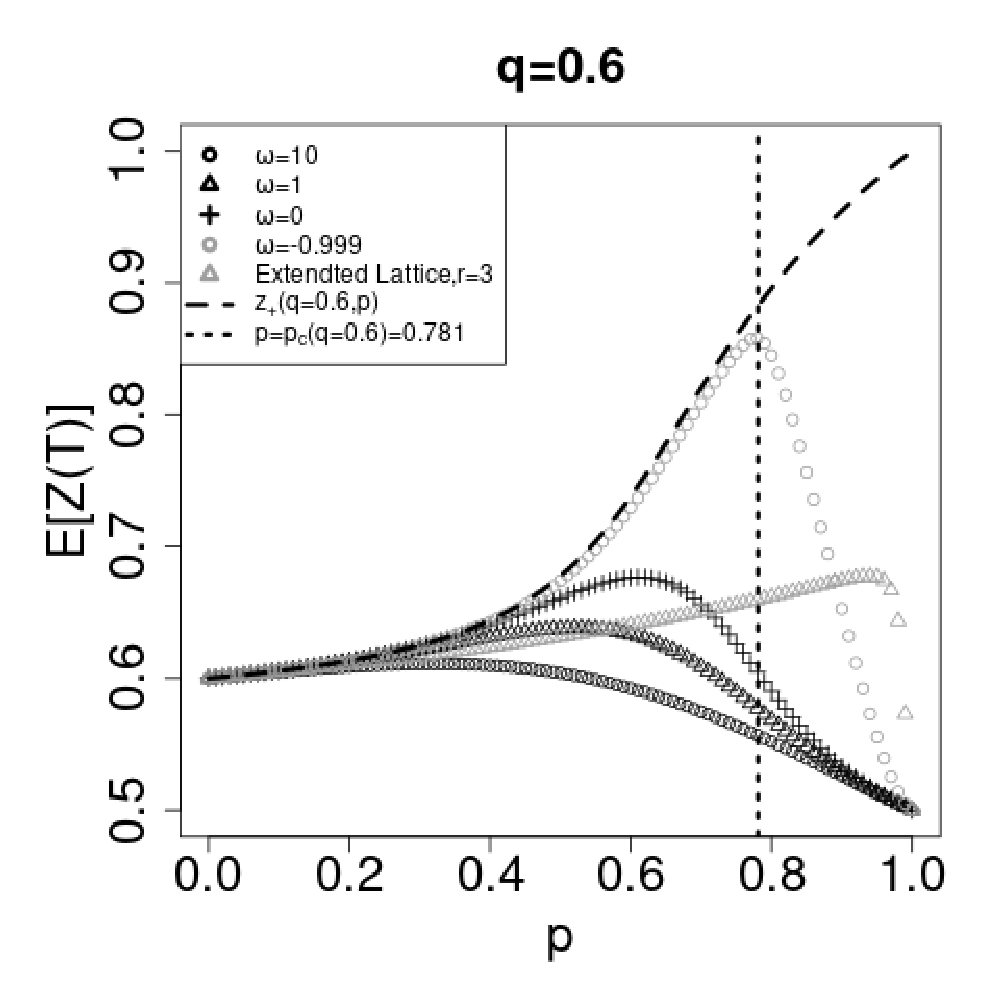}
\end{tabular}
\end{center}
\caption{
Plot $\gamma$ vs. $p$ for $q=0.6,r=3$ and $T=10^{4}$,  
$\omega=1(\mbox{black} \circ),0.0(\mbox{black} \bigtriangleup),-0.5(\mbox{gray} \circ)$
and $-0.9(\mbox{gray} \bigtriangledown)$.
The broken solid and dotted lines show the values of $p_c(q=0.6)=0.781$ and
$p_{vc}(q=0.6,\omega)$ for $\omega\in \{1,0.0\}$.
There is no super-phase for $\omega\in \{-0.5,-0.9\}$.
Right: Plot of the $z_{+}(q=0.6,p)$, the unique solution of Eq.(\ref{sc1})
for $p<p_{c}$ and the larger solution of Eq.(\ref{sc1}) for $p>p_{c}$ and
the estimated expected value of $Z(T)$ vs. $p$.
$\omega=10(\mbox{black} \bigcirc),1(\mbox{black} \bigtriangleup),0(\mbox{black}+),-0.999(\mbox{gray} \bigcirc)$ and the extended lattice
$(gray \bigtriangleup)$. The thin dotted line indicates $p=p_{c}(0.6)$.
When the extended lattice case, we  confirm there is no phase transition as $(gray \bigtriangleup)$.}
\label{fig:gamma_q06_and_EZ}
\end{figure}	

The left figure of Fig.\ref{fig:gamma_q06_and_EZ}
plots $\gamma$ vs. $p$ for $\omega\in \{1,0,-0.5,-0.9\}$.
The horizontal axis represents the ratio of herders $p$.
The vertical axis represents the speed of convergence $\gamma$.
The vertical broken line indicates the critical point, $p_{c}(q=0.6)$.
The other vertical dotted lines show $p_{vc}$ for $\omega\in \{1,0\}$,
respectively. It is clear that the phase transition
at $p=p_{c}(q=0.6)$. For $p>p_{c}$, $\gamma$ is almost zero and
the variance in $Z(t)$ becomes constant. There are two stable states.
For $p>p_{c}$, $V(Z(t))$ does not decrease to zero.
For $p<p_{c}$, a super-normal transition occurs. The positions of the
critical points are vague; evidently, the region $p$ for the
normal phase for $\omega=1$ is extremely narrow, and it considers $p_{vc}(\omega=1)=0$. For $\omega=0$, $p_{vc}\simeq 0.36$ and a plateau region was observed for $p<0.3$.
As $\omega$ decreases from $\omega=1\to -0.9$, the plateau region widens.
Near the critical point $p_{vc}$, $\gamma$ is smaller than one, and estimating $p_{vc}$ from the figure is difficult.

\begin{figure}[htbp]
\begin{center}
\begin{tabular}{cc}
\includegraphics[width=8cm]{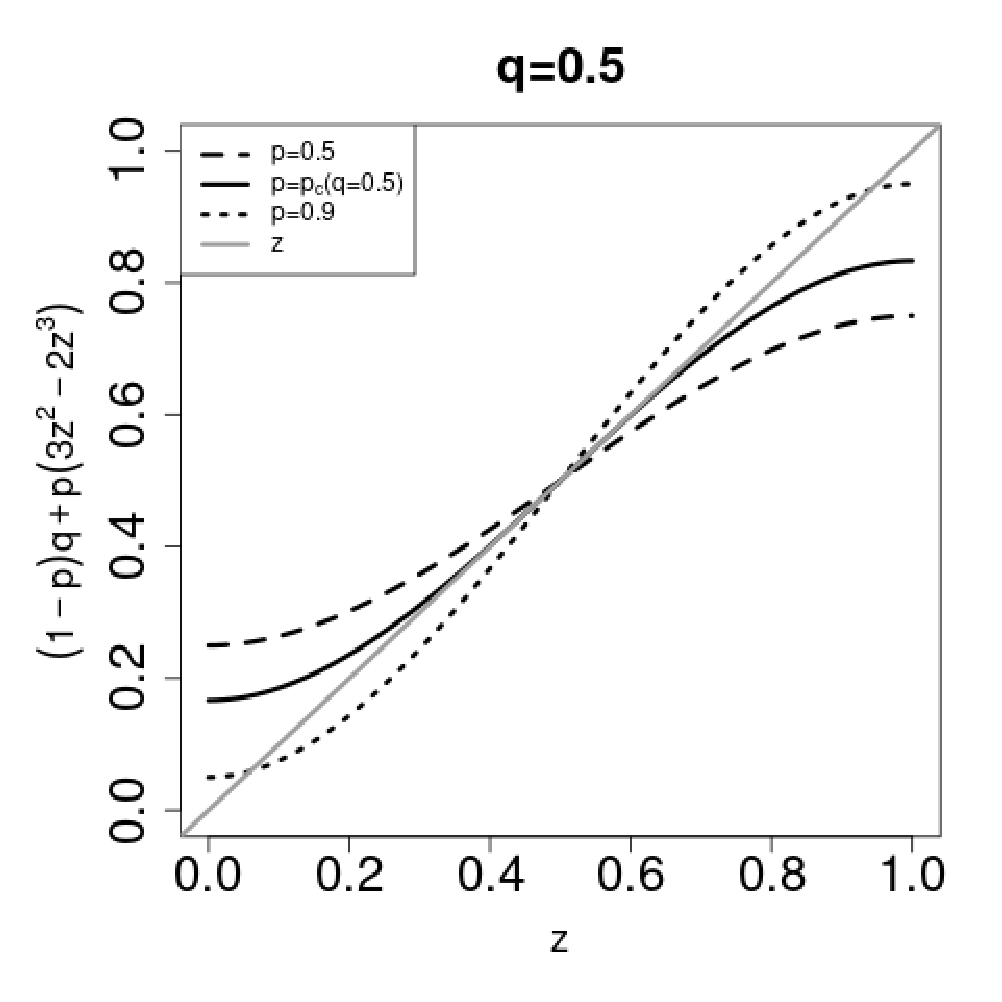} &
\includegraphics[width=8cm]{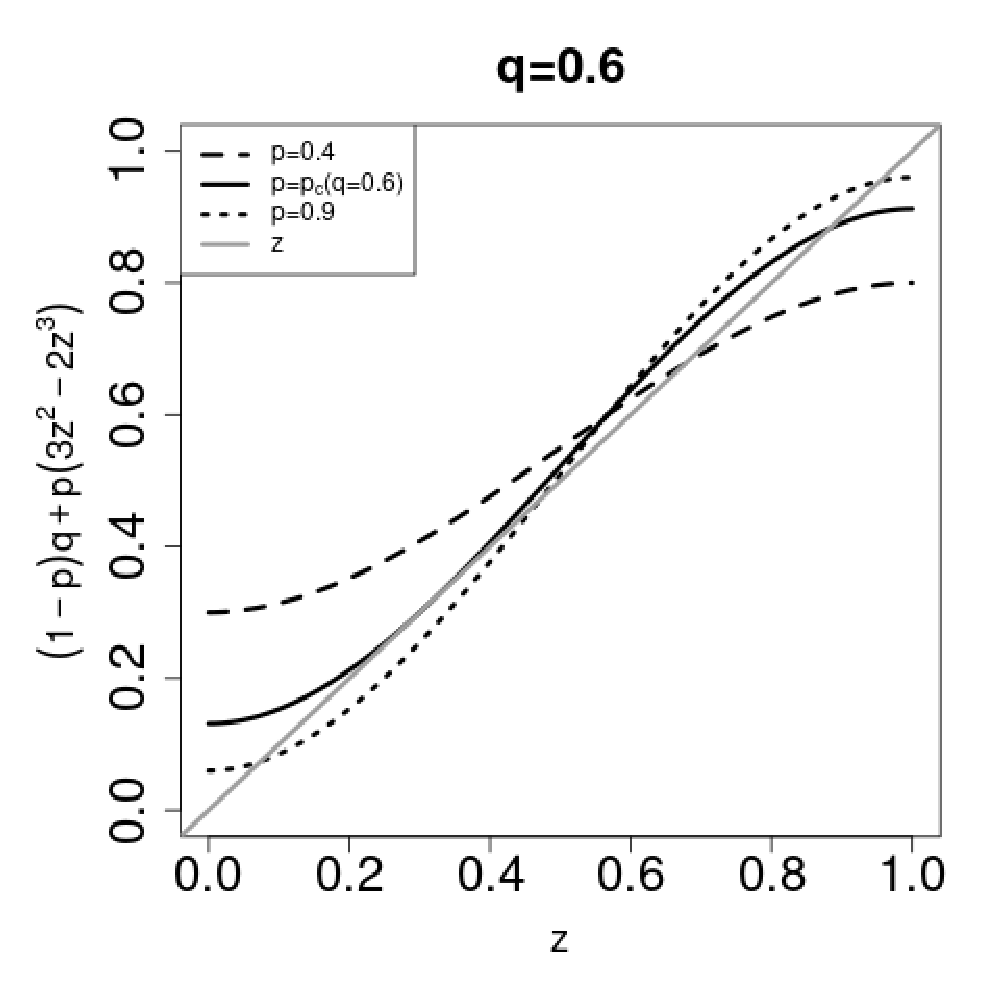}
\end{tabular}
\end{center}
\caption{Solutions of Eq.(\ref{sc2}) for the case $r=3$ and $q=0.5,0.6$.
The left figure shows the symmetric case and the right figure shows the asymmetric case.
The horizontal axis is $Z$, and the vertical axis is the RHS of Eq.(\ref{sc2}).
In both figures,  when $p<p_c$, there is only one solution.
When $p>p_c$, there are three solutions. The right solutions are $z_+$ and the left solutions are $z_-$.
$z_+$ and $z_-$ are the stable solutions.
}
\label{pvsf}
\end{figure}

In the asymmetric case, the phase transition at $p_{c}(q),q>1/2$
is discontinuous. The image on the right side of Fig.\ref{fig:gamma_q06_and_EZ}
shows the average value of $Z(T)$, which is the correct value.
The dashed curve $z_{+}$ in the  right figure  of Fig.\ref{fig:gamma_q06_and_EZ}   corresponds to the upper  slanting  solid  line and the connected dotted line in  Fig.\ref{order}.
We confirm the gap increases as $\omega$ increases.
Note that if we increase $T$,  $p_{c}$ becomes the thin dotted line.
In addition to this network model,
$\omega\in \{10,1,0,-0.999\}$, we show the results for  the 
extended lattice case. 
Because there is only one stable state for $p<p_{c}(q)$,
$Z(t)$ converges to the solution to Eq.(\ref{sc2}).
We denote this solution as $z_{+}(q,p)$ in Fig.\ref{pvsf}.
The figure shows that the convergence speed crucially
depends on $\omega$. When $\omega=10$ and $T=10^4$, as $p$ increases from 0 to 1, the average value of $Z(T)$ deviates from $z_{+}(q=0.6,p)$.
As $\omega$ decreases to -1, the departure timing is delayed, and
at $\omega=-0.999$, the average value coincided with $z_{+}(q,p)$ until
$p_{c}(q)$. 
For $p \geq p_{c}$, there are two stable solutions to Eq.(\ref{sc2}), and we denote them as $z_{+}(q,p)=2\bar{\hat{v}}_+ -1$ and $z_{-}(q,p)=2\bar{\hat{v}}_- -1$. $z_{+}(q,p)$ is a larger solution,
and the unique solution for $p<p_{c}(q)$ continuously becomes $z_{+}(q,p)$ for
$p>p_{c}(q)$. 
In the case $p< p_{c}(q)$, the expected value is the weighted average of $z_{+}(q,p)$ and $z_{-}(q,p)$. 
In the limit $T\to \infty$, $E(Z(T))$
converges to $z_{+}(q,p)$.
These  results suggest that the discontinuity in $E(Z(T))$
in the limit $T\to \infty$ at $p=p_{c}$ depends on $\omega$.

From the viewpoint of  performance, 
network $\omega\sim -1$ exhibits the best performance among all networks.
There was a phase transition; however, the gap was small.
Hence, the highest correct ratio can be confirmed for $\omega\geq-1$.
However, when there is a large hub, the gap becomes large and the performance worsens in the two-peak phase.
When the lattice case $\omega=-1$, there is no phase transition; however, the performance is worse than that of the other networks.

\section{VI. Concluding Remarks}
In this study, we examined a voting model by considering several different types of networks, including 
a random graph, the Barab\'{a}si--Albert(BA) model, and lattice networks, using one parameter $\omega$.
For a positive $\omega$, the network has hubs.
However, for negative $\omega$, the network has a limit on the number of links. 
Using $\omega$, we  continuously studied the effects of  the network.

We investigated  the phase differences between different networks as models for  voter information.
This is based on the assumption that voters obtain information from a network, including hubs.
A voting model represents how public perceptions are conveyed to voters.
Our voting model was constructed using two types of voters---herders and independents---and two candidates.
Herders vote for the majority of candidates and obtain information related to previous votes from their networks.
We examined the differences between the phases on which networks depend.

In $\omega>-1$,  we observed two types of phase transitions: 
cascade transition  and super-normal transition. 
As the number of herders increases, the model features a phase transition beyond which  a state in which most voters made the correct choice coexisted with one in which most of them were wrong.
During this transition, the distribution 
of votes changes from the one-peak phase to the two-peak phase.
This is an absorption  transition that belongs to the non-equilibrium phase transition \cite{Hin, Lu}. 
The other transition is a super-normal transition in terms of convergence speed.

In the symmetric case,  the information cascade transition is continuous  
transition. However, in the asymmetric case, it is discontinuous.
We confirmed that the transition point of the information 
cascade transition does not depend on $\omega$ for
both  symmetric and asymmetric cases.
In the symmetric case,  the information cascade transition does not depend on the network.
However, in the asymmetric case, the gap in the discontinuous phase transition depends on the network.
As $\omega$ increases, the gap increases.
In other words, the hubs affect the gap in  the  discontinuous phase transition.
In the two-peak phase, hubs affect the correct ratio.
However, in the one-peak phase, the hubs do not affect the correct ratio.

The super-normal transition also depends on the network.
As $\omega$ increases,  the transition point $p_c$ decreases.
 At $\omega=1$, the transition disappears, and  at $\omega\geq 1$,  convergence is slow for any $p$.
Hence, a phase transition is observed in $1>\omega> -1$.
This is the same as the percolation model for the network,  as shown  in Appendix E.
This is effective for both  symmetric and asymmetric cases.

For $\omega=-1$ which belongs to the lattice network, 
there was no phase transition.
Hence, the correct ratio increases as $p$ increases
because there is no phase transition.
The lattice case is summarized in Appendix C.

From the viewpoint of  performance, near the lattice case, $\omega\sim-1$ exhibits  the best performance  of the voting in all networks. 
 As the hub  size decreases,  the performance improves.
Hence, the size of hubs is the worse effects in the information cascade. 
It may be related to the network of  deep learning.
In the deep learning,  there is no hub  in the network.


Table \ref{game}  summarizes the relations between the networks and  the information cascade transition.
In this study, we considered the networks corresponding to  $\omega\neq -1$  and  the lattice with $\omega=-1$.
In the lattice case,
there was no phase transition \cite{Hisakado4}.
In  contrast, in networks with $\omega\neq -1$,
we  observe a phase transition.
In the symmetric case, it is a continuous transition with critical exponent $\beta=1/2$.
The universality class is the same as that of
the nonlinear P\'{o}lya model \cite{Hill, Pem1, Pem2}.
In the asymmetric case, the transition is discontinuous.
When all voters are refereed, the model belongs to 
  the universality class of the voter model \cite{Hin,Lu}.
In this case, the critical exponent is $\beta=1$.
This is also  an absorption  transition that belongs to the non-equilibrium phase transition.

\begin{table}[tbh]
\begin{center}
\caption{Networks and the universality class in the voting model}
\begin{tabular}{|c|c|c|c|} 
\hline
 Model & All $r=t$ &Networks $\omega\neq -1$& Lattice $\omega=-1$\\ \hline
Symmetry & $\beta=1$\cite{mori2} & $\beta=1/2$ \cite{mori1,mori3}& Oscillation \cite{Hisakado4}\\ \hline
Asymmetry& $\beta=1$ \cite{mori2}& first kind phase transition & Oscillation \cite{Hisakado4}\\ \hline
\end{tabular}
\label{game}
\end{center}
\end{table}

\begin{acknowledgments}
This work was supported by JPSJ KAKENHI[Grant No.22K03445]. 
\end{acknowledgments}

\def\thesection{Appendix \Alph{section}}

\section{Appendix A. Relation to Elephant Walk }

In this Appendix, we consider the relationship between 
the voting model used in this study and the
elephants walk model over a network \cite{ele}.
In the random graph case, and $r=1$, the model corresponds to standard elephant work.
Here, the number of referred voters is $r$.
In the elephant model, $\omega$ is the parameter  related to  the memory.
As $\omega$ increases  old memory becomes important and the old behavior is frequently remembered.
It  corresponds to   the size of hub  becoming  large.

Here, we set $p=1$ and consider the case where all voters are herders in the voting model.
We use $\hat{p}$ as the parameter instead of $p$ in the voting model.
We define $P_1^{r}(t)$ as the probability that the $t$-th voter would vote for $C_1$.
We change Eq.(\ref{ge}) to
\begin{equation}
P_{1}^{r}(t)=\left\{ 
\begin{array}{ll}
\hat{p}&:c_1^r(t)>r/2, \nonumber \\
\hat{p}/2&;c_1^r(t)=r/2, \nonumber \\
(1-\hat{p}) &:c_1^r(t)<r/2 \nonumber.
\end{array} 
\right \}
\label {ge2}
\end{equation} 
 With probability $\hat{p}$, the voter decides  on the basis of the 
referred voters.
That is, the difference between $p$ and $\hat{p}$ is the introduction of noise.
In the voting model, the voter behaves as  a herder.
In other words, the voter behaves in accordance with the information obtained.
In  contrast, in the elephant walk  model, the voter behaves 
against the referred information with probability 
$1-\hat{p}$.
In the case $\hat{p}<1/2$, the probability that the voter behaves against the referred information is larger than the voter behaves according to the obtained information.
In the case $r=1$, the model becomes an elephant walk model. 
When the analog herder case corresponds to the elephant model, the phase diagram becomes the same as that of the $r=1$ digital herder case.
A voter refers to only one  previous voter.
Analog herders vote for each candidate with probabilities proportional to the candidates' votes.
The initial condition of  the elephant model is 
$
P_{1}^{r}(1)=
\hat{q}
$ and includes an asymmetric case.
Here, we consider the symmetric case, $\hat{q}=1/2$.

We can write the evolution of connectivity as
\begin{eqnarray}
& &g_1(t)=\hat{k} \rightarrow \hat{k}+i:
\nonumber \\
& r/2\leq j\leq r, i=\omega j+r& P_{\hat{k},t}(i)={}_{r}C_{j}\hat{Z}^{j}(1-\hat{Z})^{r-j}\hat{p},
\nonumber \\
& 0\leq j< r/2, i=\omega j+r& P_{\hat{k},t}(i)={}_{r}C_{j}\hat{Z}^{j}(1-\hat{Z})^{r-j}(1-\hat{p}),
\nonumber \\
& r/2\leq j\leq r, i=\omega j& P_{\hat{k},t}(i)={}_{r}C_{j}\hat{Z}^{j}(1-\hat{Z})^{r-j}(1-\hat{p}),
\nonumber \\
& 0\leq j< r/2, i=\omega j& P_{\hat{k},t}(i)={}_{r}C_{j}\hat{Z}^{j}(1-\hat{Z})^{r-j}\hat{p},
\nonumber \\
\end{eqnarray}
and
$P_{\hat{k},t}(i)$ is the probability of the transition  from $\hat{k}$ to $\hat{k}+i$ at $t$.

Approaching the continuous limit, as in Section IV, we can obtain the stochastic partial differential equation: 
\begin{eqnarray}
\textrm{d}\hat{X}_\tau&=&[\frac{\omega}{(1+\omega)}\frac{\hat{X}_{\tau}}{(\tau-r+1)}
\nonumber \\
& &
+\frac{r(2p-1)}{1+\omega}\{ \frac{2\cdot(2n+1)!}{(n!)^2}\int_0^{\frac{1}{2}
+\frac{\hat{X}_\tau}{2r(\tau-r+1)}}x^n(1-x)^ndx-1\} ]\textrm{d}\tau+\sqrt{\epsilon}{\rm d}B.
\label{ew}
\end{eqnarray}
The derivation of Eq.(\ref{ew})     is  the extension of    the previous studies and see  \cite{Hisakado5} in detail. 
When we set $2\hat{p}-1=p$, Eq.(\ref{ew}) becomes Eq.(\ref{ito0n}).

For $r=1$, the equation becomes
\begin{equation}
\textrm{d}\hat{X}_\tau=[\frac{2p-1+\omega}{1+\omega}\frac{\hat{X}_\tau}{\tau}]\textrm{d}\tau+\sqrt{\epsilon}{\rm d}B.
\label{ew1}
\end{equation}

We can assume that the stationary solution is
\begin{equation}
\hat{X}_\infty=r\bar{\hat{v}}\tau+r(1-p)(2q-1)\tau,
\label{hn}
\end{equation}
where $\bar{\hat{v}}$ denotes a constant.
As Eq.(\ref{xz}) and $0\leq\hat{Z}\leq1$, we obtain
\begin{equation}
-1\leq\bar{\hat{v}}+(1-p)(2q-1)\leq1.
\end{equation}
Substituting Eq.(\ref{hn}) into Eq.(\ref{ew}), we obtain
\begin{equation}
\bar{\hat{v}}=(2p-1)\{\frac{2\cdot(2n+1)!}{(n!)^2}\int_0^{\frac{1}{2}+
\frac{\bar{\hat{v}}}{2}} x^n(1-x)^ndx -1\}.
\label{i3}
\end{equation}
This equation is self-consistent. 
The equation does not depend on parameter $\omega$.

We expand $\hat{X}_\tau$ around solution $r\bar{\hat{v}}\tau+r(1-p)(2q-1)\tau$, 
\begin{equation}
\hat{X}_\tau=r\bar{\hat{v}}\tau+r\hat{W}_\tau.
\label{w4}
\end{equation}
We set $\hat{X}_\tau\gg \hat{W}_\tau$.
This result indicates that $\tau\gg 1$.
We rewrite Eq.(\ref{ew}), using Eq.(\ref{w4}), and obtain the following:
\begin{eqnarray}
\textrm{d}\hat{W}_\tau
&=&[\frac{\omega+(2p-1)A}{1+\omega}]\frac{\hat{W}_\tau}{\tau}\textrm{d}\tau+\sqrt{\epsilon}{\rm d}B,
\label{ito44}
\end{eqnarray}
where 
\begin{equation}
A=\frac{(2n+1)!}{(n!)^2\cdot 2^{2n}}(1-\bar{\hat{v}}^2)^n.
\end{equation}

\begin{figure}[h]
\includegraphics[width=110mm]{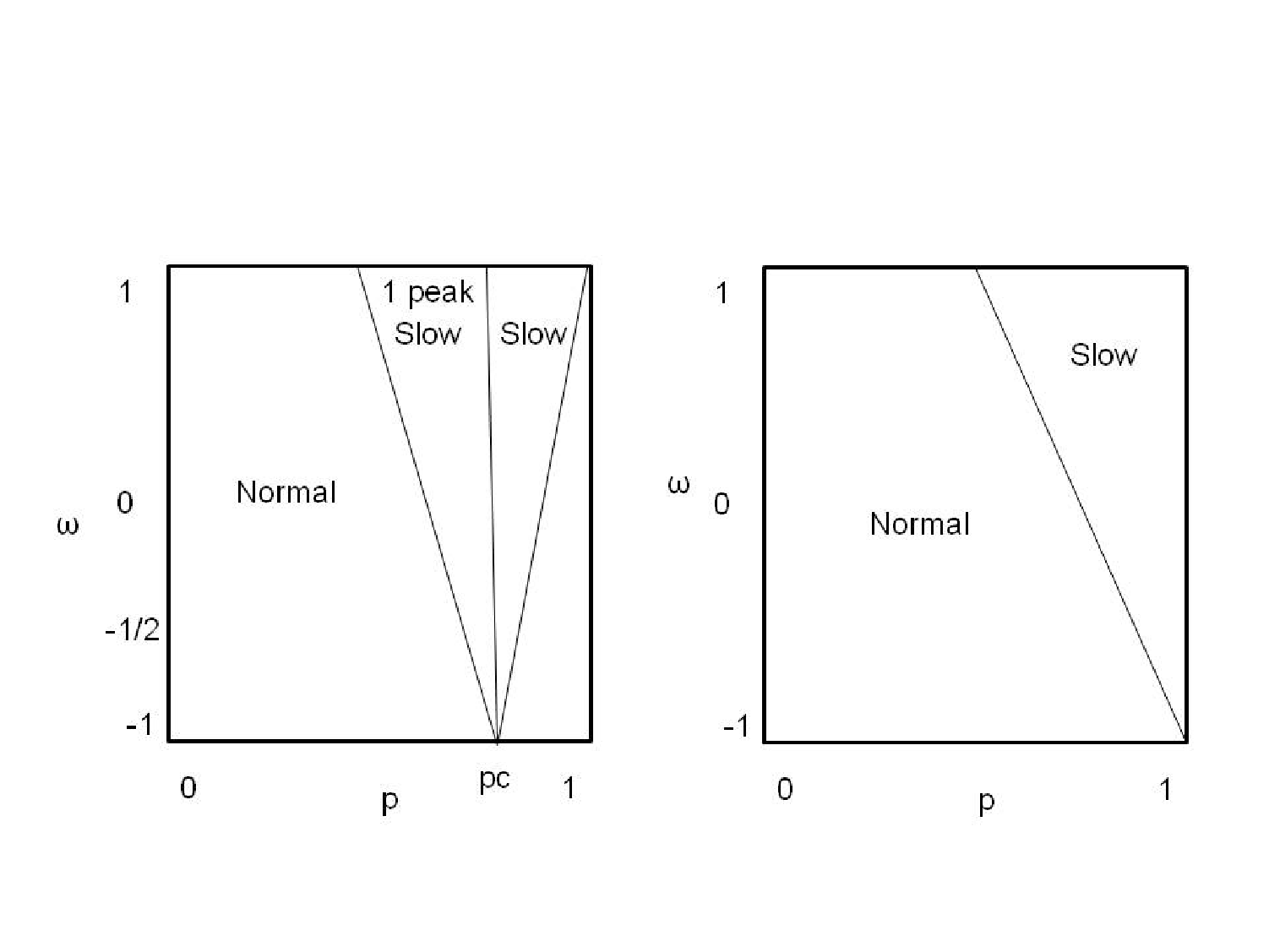}
\caption{Phase diagram of elephant walk for $r=1$ and $r=3$ cases for the symmetric case $q=1/2$. In the case $r=3$, $p_c$ is 2/3. }
\label{phase2}
\end{figure}

The phase diagram of the elephant walk is shown in Fig.\ref{phase2}. 
We can map from Fig.\ref{phase2} to Fig.\ref{phase}, if we set $2\hat{p}-1=p$.
The case $r=1$ corresponds to  the elephant walk.
We can confirm a super-normal transition 
 of the basic elephant work model at $\hat{p}=3/4$ \cite{ele}.

\subsection{B. How to create the network}

\begin{figure}[h]
\includegraphics[width=110mm]{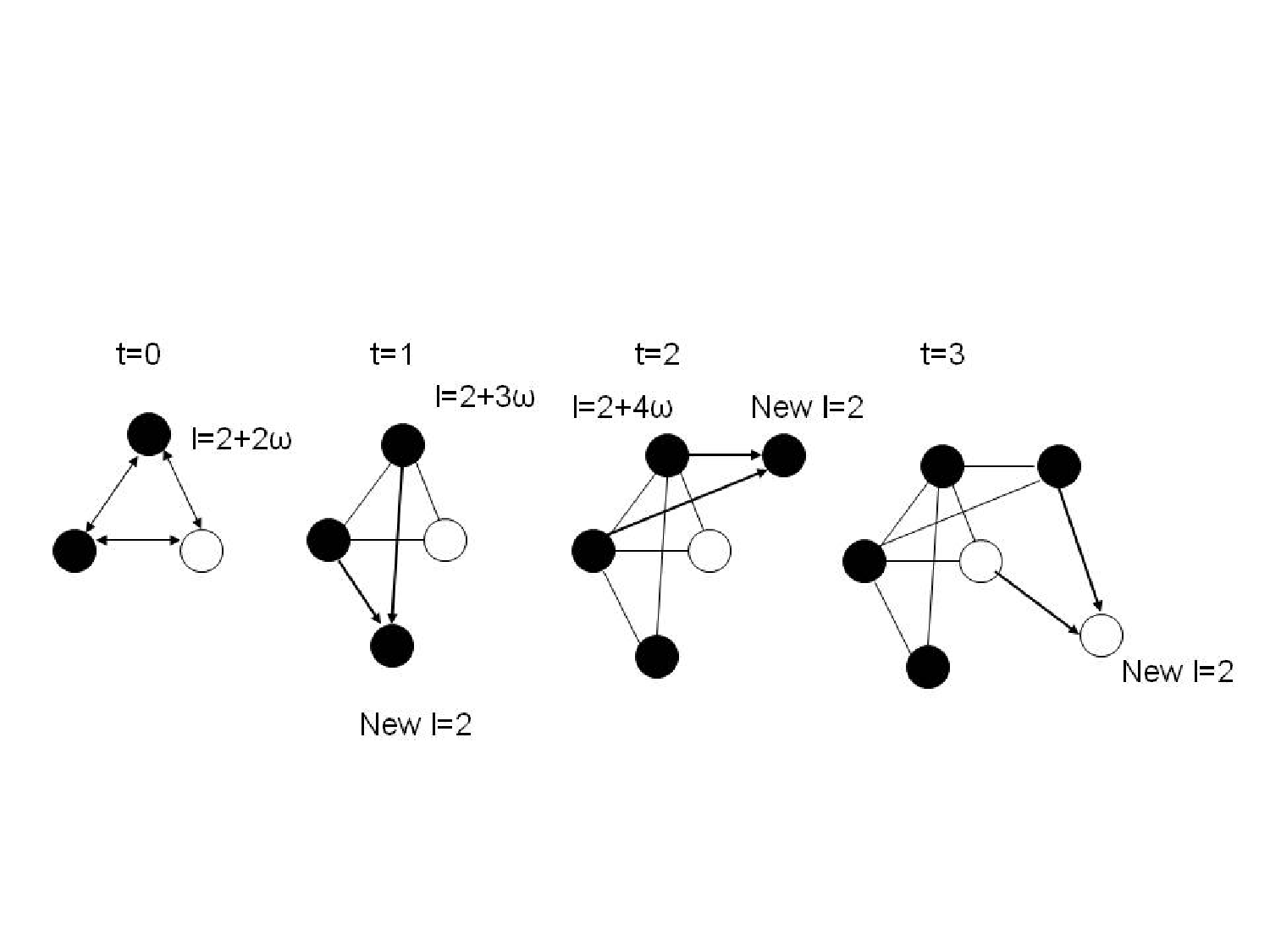}
\caption{
A few   steps of the creation of the networks are shown.  The initial condition $t=0$ is the perfect graph of 3 nodes,  and $r=2$. The bold directed edge represents the new links from the referred  voters to the new voter. It shows the flow of  information. The white (black) dot is a voter who voted for  candidate $C_0 (C_1)$.  }
\label{IC}
\end{figure}

In this Appendix, we explain  how to create  networks.
We show   the initial   few steps of the  network
in Fig.\ref{IC}.
We begin  with the creation of the directed network by a process similar
to the BA preferential attachment model. 
In this network, each node $i$ has in-degree $k_i^{IN}$ and out-degree $k_i^{OUT}$. The popularity of each node is
defined as $l_i=k_i^{IN}+\omega k_i^{OUT}$. 

The initial condition of an example, $t=0$ is the perfect graph,  and $r=2$.
In the BA model case, the perfect graph is often used as the initial condition.
Here we use the perfect graph, but it is not a  necessity condition.
We begin the perfect graph with 3 nodes and
each node has  two incoming  arrows  and two outgoing arrows.
The popularity of   each node is $2+ 2\omega$ at $t=0$.
In this study, we  also use the perfect graph for the numerical simulations.
The initial condition does not affects the conclusion of this article.
The initial nodes are  independent voters and 
decide their  opinion independently.
After that,  the   herders and the independents  appear.
The white (black) dot is a voter who voted for candidate $C_0 (C_1)$. 

 At step $t=1$ the fourth  node is attached  by a directed link going from 
existing nodes   to a newly created node. 
The arrow indicates  the flow of the information.
When the new node is a  herder,  information is referred for  voting.
When the new node is   independent, the information is not used for  voting.
The popularity of the fourth node is $2$ because of the  two incoming arrow, while the
popularity of the initial  node  that  are referred  is $2+3\omega$.
When $\omega$ is negative,
 popularity  decreases as  time goes by, and
the  node that  has 0 or negative popularity 
will never  be selected.

\section{Appendix C. Lattice Case}
In this section, we summarize the extended lattice case \cite{Hisakado4}.
Here, we consider a random walk between the two states $c_1^{r} (t)/r>1/2$ (good equilibrium) and $c_0^{r}(t)/r>1/2$ (bad equilibrium). 
We define the hopping probability from state $c_0^r (t)/r>1/2$ to $c_1^r (t)/r>1/2$ as $a$ and
that from  state $c_1^{r} (t)/r>1/2$ to $c_0^{r} (t)/r>1/2$ as $b$. $a$ and $b$ are not functions of $t$. When $t>r$, the transition matrix $\hat{A}$ of the random walk is
\begin{equation}
\hat{A}=\left (
\begin{array}{ll}
1-a& a \nonumber \\
b & 1-b
\end{array}
\right ).
\label{proo}
\end{equation}
The random walk of the two states is defined as the  transition matrix $\hat{A}$ when $t>r$. If $r > t$, voters can view $t$ previous votes for each candidate.

If consecutive independent voters choose candidate $C_1(C_0)$ when $c_0^{r} (t)/r>1/2 (c_1^{r} (t)/r>1/2) $, the state changes from $c_0 ^{r} (t)/r>1/2 (c_1^{r} (t)/r>1/2) $ to $c_1^{r} (t)/r>1/2 (c_0^r (t)/r>1/2) $. Thus, independent voters act as switches for hopping. When independent voters who vote $C_1(C_0)$ are the majority, the state hops from $c_0^r (t)/r>1/2 (c_1^r (t)/r>1/2) $ to $c_1^r (t)/r>1/2 (c_0^r (t)/r>1/2)$.
Hence,  the hopping rates $a$ and $b$ are estimated as follows:
\begin{eqnarray}
a&=&\pi[(1-p)q]=\frac{(2n+1)!}{(n!)^2}\int_0^{(1-p)q}x^n(1-x)^ndx\sim (1-p)^{\frac{r+1}{2}}q^{\frac{r+1}{2}}, \nonumber \\
b&=&
\pi[(1-p)(1-q)]
\sim (1-p)^{\frac{r+1}{2}}(1-q)^{\frac{r+1}{2}},
\label{pi1}
\end{eqnarray}
where the approximations were $p\sim 1$. In the case where $r=1$, $a=(1-p)q$ and $b=(1-p)(1-q)$. We obtained a solution identical to that in \cite{Hisakado4}.

In the finite $r$ case,  the hopping rates $a$ and $b$ do not decrease as $t$ increases, and the state oscillates between good and bad equilibria.
Hence, the distribution of $Z(t)$ becomes normal and there is no phase transition. The voting rate converges to $(1-p)q+pa/(a+b)\sim(1-p)q+pq^{\frac{r+1}{2}}/(q^{\frac{r+1}{2}}+(1-q)^{\frac{r+1}{2}})$.  The first term is the number of votes by independent voters, and the second term is the number of votes by digital herders. The herders' votes oscillate between good and bad equilibria in Eq.(\ref{i3}). As $r$ increases, the stay in good equilibrium becomes longer. The ratio of stay in a good equilibrium to stay in a bad equilibrium is $a/b\sim (\frac{q}{1-q})^{r+1/2}$.

\section{Appendix D. Universal Function for Symmetric Model}
In this section, we describe the calculation of the universal function for the symmetric case \cite{mori1,mori2,mori3}.
Here, we consider  the correlation function $C(t)$.
The asymptotic behavior $C(t)$
is described by the scaling form for the symmetric case as
 \begin{equation}
   C(t)\propto(\ln t)^{-1/2}g(\ln t/\xi),
 \end{equation}
 where $g(x)$ is the universal fun cation and
 $\xi$ denotes the correlation length \cite{mori1,mori2}.
 $\xi$ characterizes 
 length scale of the correlation function $C(t)\propto e^{-t/\xi}$.

We used the new random variable $\hat{Y}_{\tau} = \hat{X}_{\tau}/(\tau-r+1)$,
\begin{equation}
 d\hat{Y}_{\tau} = \frac{\varphi(\hat{Y}_{\tau})}{\tau-r+1} d\tau
 + \frac{\sqrt{\varepsilon}}{\tau-r+1}{\rm d}B,
\label{itouf}
\end{equation}
where
\begin{eqnarray}
 \varphi(y) &= &f(y) - y
 = \frac{2r}{1+\omega}\left\{
  (1-p)(q-1/2)
  - \frac{y}{2r}
  + p\rho\left(\frac{y}{2r}\right)
 \right\},
 \nonumber \\
 \rho(z) &=& \pi\left(z+\frac{1}{2}\right) - \frac{1}{2},
\end{eqnarray}
 where $\pi(z)$ is the regularized beta function, Eq.(\ref{pi}), 
$I_{z}(n+1,n+1)$ \cite{HB}.
The important property of function $\rho(z)$ is
that it is a $\tanh$-shaped function, such that
\begin{eqnarray}
  \rho\left(\pm\frac{1}{2}\right) &=& \pm\frac{1}{2},\quad
  \rho(z) + \rho(-z) = 0,
  \nonumber \\
  \rho^{\prime}(z) &= &\frac{1}{p_{0}}(1-4z^{2})^{n},
  \quad
  p_{0} \equiv \frac{(2n)!!}{(2n+1)!!}.
\end{eqnarray}
When $p\le p_{0}$, function $\varphi(y)$ has only one zero
within the range $-r\le y\le r$.
However, when $p>p_{0}$, three distinct
zeros for a sufficiently small $q$ in the same interval.
In particular,
for the $q=1/2$ case,
function $\varphi(y)$ recovers 
$Z_{2}$symmetry, $\varphi(-y) = -\varphi(y)$.
Thus, the system exhibits a continuous phase transition
at critical point $p=p_{c}$ where $p_{c}=p_{0}$.

We perturbatively solve Eq.(\ref{itouf}) under the initial condition:
$\hat{Y}_{\tau} = Y_{0}$ and
We expand the random variable $\hat{Y}_{\tau}$ to powers of $\sqrt{\varepsilon}$,
\begin{equation}
 \hat{Y}_{\tau} = \hat{Y}_{\tau}^{(0)} + \sqrt{\varepsilon}\hat{Y}_{\tau}^{(1)} + (\sqrt{\varepsilon})^{2}\hat{Y}_{\tau}^{(2)} + \cdots,
\end{equation}
and Eq.(\ref{itouf}) is solved recursively.
First, the classical solution $\hat{Y}_{\tau}$ is given by
\begin{equation}
 \int_{Y_{0}}^{\hat{Y}_{\tau}^{(0)}}\!\frac{dy}{\varphi(y)} = \log(\tau-r+1).
\end{equation}
Then, $\hat{Y}_{\tau}^{(1)}$ and $\hat{Y}_{\tau}^{(2)}$ can be expressed as
\begin{eqnarray}
 \hat{Y}_{\tau}^{(1)} &=& \varphi(\hat{Y}_{\tau}^{(0)})\!\int_{r}^{\tau}\!\frac{dB_{s}}{\varphi(\hat{Y}_{s}^{(0)})(s-r+1)},
 \nonumber \\
 \hat{Y}_{\tau}^{(2)} &=& \frac{\varphi(\hat{Y}_{\tau}^{(0)})}{2}\int_{r}^{\tau}\!\frac{\varphi^{\prime\prime}(\hat{Y}_{s}^{(0)})(\hat{Y}_{s}^{(1)})^{2}}{\varphi(\hat{Y}_{s}^{(0)})(s-r+1)}\,ds.
\end{eqnarray}
Notably, the expectation values are given by
\begin{eqnarray}
 \label{eq:EY}
 \mathrm{E}[\hat{Y}_{\tau}^{(1)}] &= &0,
 \\
 \mathrm{E}[(\hat{Y}_{\tau}^{(1)})^{2}] &=& \varphi(\hat{Y}_{\tau}^{(0)})^{2}\int_{r}^{\tau}\!\frac{ds}{\varphi(\hat{Y}_{s}^{(0)})^{2}(s-r+1)^{2}},
 \\
 \mathrm{E}[\hat{Y}_{\tau}^{(2)}] &=& \frac{\varphi(\hat{Y}_{\tau}^{(0)})}{2}\int_{r}^{\tau}\!\frac{\varphi^{\prime\prime}(\hat{Y}_{s}^{(0)})E[(\hat{Y}_{s}^{(1)})^{2}]}{\varphi(\hat{Y}_{s}^{(0)})(s-r+1)}\,ds.
\end{eqnarray}

Let $Y^{\pm}_{\tau}$ denote the solution to Eq.(\ref{itouf}) with the initial conditions
$Y_{0}=\pm r$.
Subsequently, the autocorrelation function is given by
\begin{equation}
 C_{X}(\tau) = c\left\{\mathrm{E}[\hat{Y}_{\tau}^{+}+\varphi(\hat{Y}_{\tau}^{+})]
  - \mathrm{E}[\hat{Y}_{\tau}^{-}+\varphi(\hat{Y}_{\tau}^{-})]
 \right\},
\end{equation}
where $c$ denotes a constant.
As seen from Eq.(\ref{eq:EY}), the long-term behavior of $C_{X}(\tau)$
is governed by the classical solution $Y^{(0)\pm}(\tau)$
because $Y^{(0)\pm}(\tau)$ converges to zero for $\varphi(y)$,
Thus, we have
\begin{equation}
 C_{X}(\tau) \sim \frac{c\omega}{1+\omega}(Y^{(0)+}(\tau) - Y^{(0)-}(\tau))
 + O((\sqrt{\varepsilon})^{2}).
\end{equation}

Now, we concentrate on the $q=1/2$ symmetric model, $p\sim p_{c}$, and $\tau\sim\infty$ case.
Subsequently, the classical solution reads as follows:
\begin{equation}
 Y^{(0)\pm}(\tau)
 \simeq \pm c^{\prime}\sqrt{\frac{p^{\prime}}{1-(1-4p^{\prime})(\tau-r+1)^{2\varphi^{\prime}(0)}}},
\end{equation}
where
\begin{equation}
 p^{\prime} = \frac{3r^{2}((p_{c}/p)-1)}{n}.
\end{equation}
Therefore, we obtain the universal function:
\begin{equation}
  g(x) = \lim_{\tau\rightarrow\infty}\frac{C_{X}(\tau|p)}{C_{X}(\tau|p_{c})}
  = \left\{
    \begin{array}{@{\,}ll}
      \sqrt{\frac{x}{1-e^{-x}}}, &
      x = \frac{2(p/p_{c}-1)}{1+\omega}\log\tau,\quad
      p\rightarrow p_{c}+0,
      \\
      \sqrt{\frac{2x}{e^{2x}-1}}, &
      x = \frac{1-p/p_{c}}{1+\omega}\log\tau,\quad
      p\rightarrow p_{c}-0.
    \end{array}
  \right.
\end{equation}

\section{Appendix E. Molloy--Reed Condition and Percolation on Networks}

In this Appendix, we consider the Molly and Reed conditions of the networks.
Using the Molly and Reed conditions, we calculate the phase transition point.
First, we calculate the second momentum of  the degree distribution.

\subsection{A. When $\omega>0$}

Cumulative degree distribution of the network 
as
\begin{eqnarray}
 P[k_i(t)<k]&=&P\left[(\frac{r}{\omega})^{(1+\omega)/\omega}
  \frac{t}{(k+r\frac{1+\omega}{\omega})^{(1+\omega)/\omega}}<t_i\right],
\nonumber \\
&=&
\frac{1}{r+t}\left[t-(\frac{r}{\omega})^{(1+\omega)/\omega}
\frac{t}{(k+r\frac{1-\omega}{\omega})^{(1+\omega)/\omega}}\right] \nonumber,
\end{eqnarray}
where $k_i$ denotes the number of links of node $i$.
Notably, $k_i\geq r$ because node $i$ has $r$ initial links.
Hence, in the limit $t\rightarrow \infty$ the degree distribution is
\begin{equation}
p(k)=\frac{\partial P[k_i(t)<k]}{\partial k} 
= A_B\frac{1+\omega}{\omega}(k+r\frac{1-\omega}{\omega})^{-2-1/\omega},
\end{equation}
where $A_B=(r/\omega)^{1+1/\omega}$.
We can confirm the normalization
\begin{equation}
\int_{r}^{\infty}p(k)dk=A_B(r/\omega)^{-1-1/\omega}=1.
\end{equation}  
The first moment was calculated as follows:
\begin{equation}
<k>=\int_{r}^{\infty}k p(k)dk=r+r=2r.
\end{equation}
The second moment was obtained.
Here, when $0<\omega<1$,
\begin{equation}
<k^2>=\int_{r}^{\infty}k^2 p(k)dk=3r^2+\frac{2r^2}{1-\omega}.
\label{2b}
\end{equation}
When $\omega> 1$, we obtain
\begin{equation}
<k^2>= \int_{r}^{\infty}k^2 p(k)dk=3r^2+\frac{2r^2}{1-\omega}+ 2\frac{\omega^2}{\omega-1}(\frac{r}{\omega})^{1+1/\omega}(k_{Max}+r\frac{1-\omega}{\omega})^{1-1/\omega},
\label{2nd}
\end{equation}
where $k_{Max}$ is the maximum number of links and goes to infinity.
Hence, the second moment is finite when $1>\omega >0$ and
infinity, when $\omega>1$.

Hence, under the condition $1>\omega > 0$, the Molloy--Reed condition is
\begin{equation}
\frac{<k^2>}{<k>}=\frac{3r}{2}+\frac{r}{1-\omega}.
\label{Mb}
\end{equation}
When $\omega>1$, $<k^2>\rightarrow \infty$.

\subsection{B. When $\omega=0$}

Cumulative degree distribution of the network 
as
\[
P[k_i(t)<k]=P[t_i>t e^{-\frac{k}{r}}]=\frac{1}{r+t}
(t-t e^{-\frac{k}{r}}).
\]
The degree distribution in the limit $t\rightarrow \infty$ is:
\begin{equation}
p(k)= A_R e^{-\frac{k}{r}},
\end{equation}
where $A_R=e/r$ to satisfy the condition $\int_r^{\infty}p(k)dk=1 $.

The first moment was calculated as follows:
\begin{equation}
<k>=\int_{r}^{\infty}k p(k)dk=2r.
\end{equation}
Similarly, we can obtain the second moment as
\begin{equation}
<k^2>=\int_{r}^{\infty}k^2 p(k)dk=5r^2.
\label{2r}
\end{equation}
Then, the Molloy--Reed condition is
\begin{equation}
\frac{<k^2>}{<k>}=5r/2.
\label{Mr}
\end{equation}

\subsection{C. When $-1<\omega<0$}

Cumulative degree distribution of the network is
\begin{eqnarray}
P[k_i(t)<k]&=&
P\left[(\frac{|\omega|}{r})^{\frac{1-|\omega|}{|\omega|}}
 (\frac{1+|\omega|}{|\omega|}r-k)^{\frac{1-|\omega|}{|\omega|}}t>t_i\right],
\nonumber \\
&=&
\frac{1}{r+t}\left(\frac{|\omega|}{r}\right)^{\frac{1-|\omega|}{|\omega|}}
\left(\frac{1+|\omega|}{|\omega|}r-k\right)^{\frac{1-|\omega|}{|\omega|}}t. 
\end{eqnarray}

We can obtain the degree distribution in the limit $t\rightarrow \infty$:
\begin{equation}
p(k)=A_F\frac{1-|\omega|}{|\omega|}(\frac{1+|\omega|}{|\omega|}r-k)^{1/|\omega|-2},
\end{equation}
where $A_F=(r/\omega)^{1-1/|\omega|}$ to satisfy the 
condition $\int_{r}^{r+r/|\omega|}p(k)dk=1$.

The first moment was calculated as follows:
\begin{equation}
<k>=\int_{r}^{\infty}k p(k)dk=r+r=2r.
\end{equation}
Similarly, we obtain the second moment.
\begin{equation}
<k^2>=\int_{r}^{\infty}k^2 p(k)dk=3r^2+\frac{2r^2}{1+|\omega|}.
\label{2f}
\end{equation}
Then, the Molloy--Reed condition is
\begin{equation}
\frac{<k^2>}{<k>}=\frac{3r}{2}+\frac{r}{1+|\omega|}.
\label{Mf}
\end{equation}

\subsection{D. Phase transition of percolation model}
We can combine Eq.(\ref{2b}), Eq.(\ref{2r}), and Eq.(\ref{2f}).
\begin{equation}
<k^2>=3r^2+\frac{2r^2}{1-\omega},
\end{equation}
where $1>\omega>-1 $.
We can combine Eqs.(\ref{Mb}), (\ref{Mr}), and (\ref{Mf})
\begin{equation}
\frac{<k^2>}{<k>}=\frac{3r}{2}+\frac{r}{1-\omega},
\end{equation}
where $1>\omega>-1 $.
When $\omega\geq 1$, $<k^2>/<k>\rightarrow \infty$.

\begin{figure}[h]
\includegraphics[width=110mm]{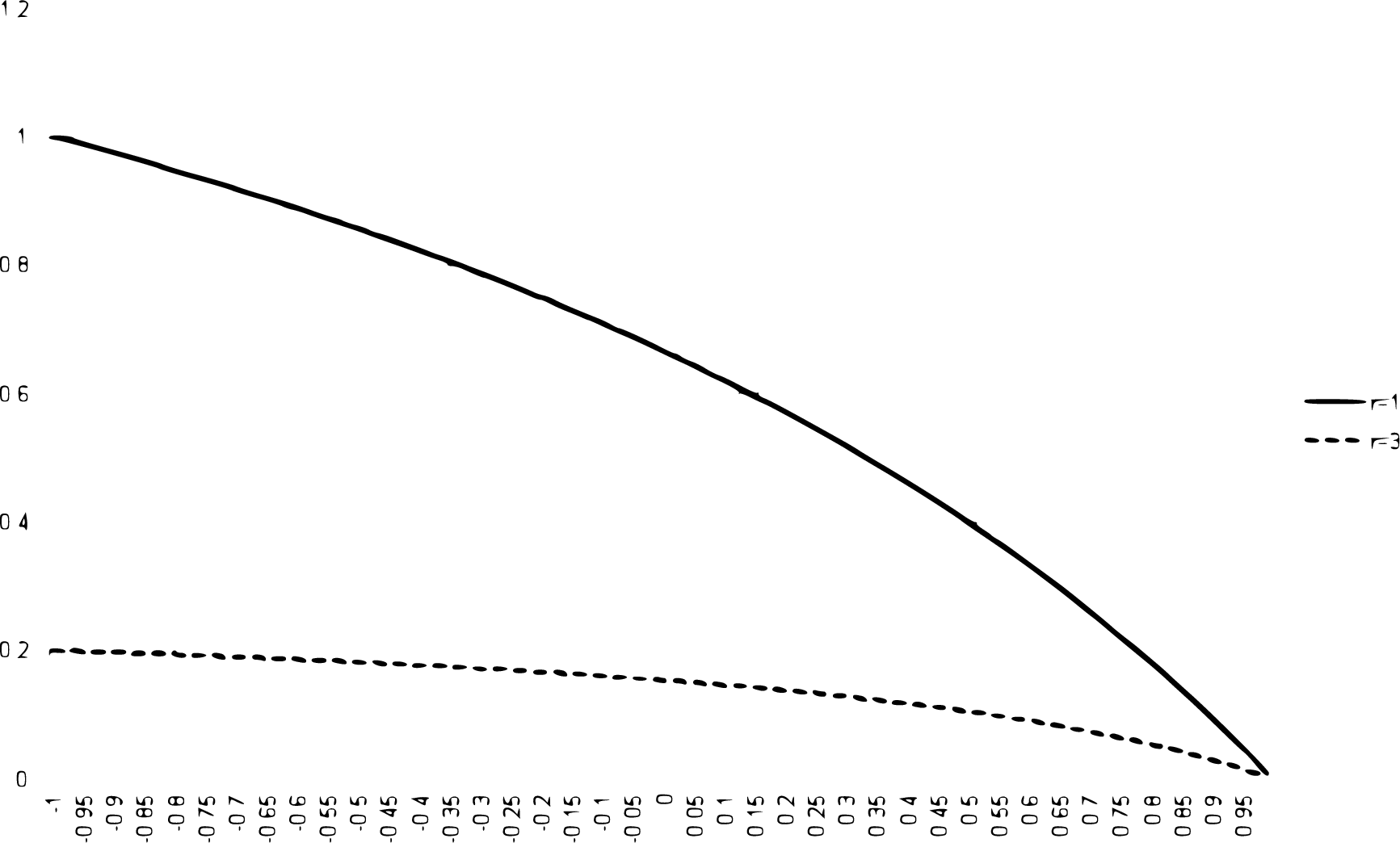}
\caption{The transition point $q_c$. The horizontal axis is $\omega$ and the vertical axis is $q_c$ There is the discontinuity in $\omega=-1$ without $r=1$. The phase transition is observed in $1>\omega>-1$.}
\label{qc}
\end{figure}

The critical probability for the percolation model is
\begin{equation}
  q_c=1/[\frac{3r}{2}-1+\frac{r}{1-\omega}],
\end{equation}
where $1>\omega>-1$.
In the case of the limit $\omega\rightarrow 1$, we can obtain $q_c=0$.
Then, $\omega\geq 1$, there is no phase transition.
In the limit $\omega=-1$, we obtain $q_c=1/(2r-1)$.
In contrast, $\omega=-1$, the network is a lattice and
$q_c=1$.
When $r=1$, $q_c$ is continuous, 
$r\neq 1$, $q_c$ is
 discontinuous.
When $\omega>1$, $q_c=0$ at $\omega=1$.
We show $q_c$ in Fig.\ref{qc}.
Hence, a phase transition was observed in $1 > \omega>-1$.

\section{Appendix F. Super-normal phase transition}
\label{Apb}
We  consider the stochastic differential equation,
\begin{equation}
\textrm{d}\hat{W}_\tau=
\hat{l}(\frac{ \hat{W}_\tau}{\tau})\textrm{d}\tau+\sqrt{\epsilon}{\rm d}B,
\label{itof}
\end{equation}
where $\tau\geq1$.
When $\hat{l}=(\omega+pA)/(1+\omega)$,  Eq.(\ref{itof}) corresponds to Eq.(\ref{ito4}).
Let $\sigma^2_1$ be the expected value of $\hat{W}_1$ and   $\sigma^2_1$
be the variance of $\hat{W}_1$.
If $\hat{W}_1$ is Gaussian   $(\hat{W}_1\sim\textrm{N}(\mu_1,\sigma^2_1))$ or deterministic $(\hat{W}_1\sim\delta_{\mu_1})$, the law of $\hat{W}_\tau$ ensures that the  Gaussian is in  accordance with density
\begin{equation}
p(\hat{W}_\tau)\sim
\frac{1}{\sqrt{2\pi}\sigma_\tau}\textrm{e}^{-(\hat{W}_\tau-\mu_\tau)^2/2\sigma_\tau^2},
\end{equation}
where $p(\hat{W}_\tau)$  is the distribution of $\hat{W}_\tau $, $\mu_\tau=\textrm{E}(\hat{W}_\tau)$ is the expected value of $\hat{W}_\tau$, and 
$\sigma^2_\tau\equiv \nu_\tau$ is its variance $\hat{W}_\tau$.
If $\Phi_\tau(\xi)=\log(\textrm{e}^{\textrm{i}\xi \hat{W}_\tau})$
 is the logarithm of the characteristic function of the law of $\hat{W}_\tau$, we have
\begin{equation}
\partial_\tau
\Phi_\tau(\xi)
=\frac{\hat{l}}{\tau}\xi\partial_\xi\Phi_\tau(\xi)
-\frac{\epsilon}{2}\xi^2,
\end{equation}
and
\begin{equation}
\Phi_\tau(\xi)=\textrm{i}\xi \mu_\tau-\frac{\xi^2}{2}\nu_\tau.
\end{equation}
Identifying
the real and imaginary parts of $\Phi_\tau(\xi)$, we 
obtain the dynamics of  $\mu_\tau$ as
\begin{equation}
\dot{\mu}_\tau=\frac{\hat{l}}{\tau}\mu_\tau.
\end{equation}
The solution for $\mu_\tau$ is
\begin{equation}
\mu_\tau=\mu_1\tau^{\hat{l}}.
\end{equation}
The dynamics of $\nu_\tau$ are  given by the Riccati equation
\begin{equation}
\dot{\nu}_\tau=\frac{2\hat{l}}{\tau}\nu_\tau+\epsilon.
\label{gp}
\end{equation}
If $\hat{l}\neq 1/2$, we get
\begin{equation}
\nu_\tau=
\nu_1\tau^{2\hat{l}}+\frac{\epsilon}{1-2\hat{l}}(\tau-\tau^{2\hat{l}}).
\end{equation}
If $\hat{l}=1/2$, we get
\begin{equation}
\nu_\tau=\nu_1\tau+\epsilon\tau\textrm{log}\tau.
\end{equation}
We can  summarize the temporal behavior of the variance as 
\begin{equation}
\nu_\tau\propto\frac{\epsilon}{1-2\hat{l}}\tau\hspace{1cm}\textrm{if}\hspace{0.5cm}\hat{l}<\frac{1}{2},
\end{equation}
\begin{equation}
\nu_\tau\propto(\nu_1+\frac{\epsilon}{2\hat{l}-1})\tau^{2\hat{l}}\hspace{1cm}\textrm{if}\hspace{0.5cm}\hat{l}>\frac{1}{2},
\end{equation}
\begin{equation}
\nu_\tau\propto\epsilon\tau\textrm{log}(\tau)\hspace{1cm}\textrm{if}\hspace{0.5cm}\hat{l}=\frac{1}{2}.
\end{equation}
Here, we introduce rescaled variables
\[
\tilde{\nu}_\tau\equiv\frac{\nu_\tau}{\tau^2}.
\]
The solution for $\tilde{\nu}_\tau$ is
\begin{equation}
\tilde{\nu}_\tau\propto\frac{\epsilon}{1-2\hat{l}}\tau^{-1}\hspace{1cm}\textrm{if}
\hspace{0.5cm}\hat{l}<\frac{1}{2},
\end{equation}
\begin{equation}
\tilde{\nu}_\tau\propto(\nu_1+\frac{\epsilon}{2\hat{l}-1})\tau^{2\hat{l}-2}\hspace{1cm}
\textrm{if}\hspace{0.5cm}\hat{l}>\frac{1}{2},
\nonumber \\
\end{equation}
\begin{equation}
\tilde{\nu}_\tau\propto\epsilon\frac{\textrm{log}(\tau)}{\tau}
\hspace{1cm}\textrm{if}\hspace{0.5cm}\hat{l}=\frac{1}{2}.
\end{equation}
This model has three phases. If $\hat{l}>1/2$ or $\hat{l}=1/2$,
    $\hat{W}_\tau/\tau$ converges  slower than  in  a binomial distribution.
These phases  are the super diffusion phases.  
If  $0<\hat{l}<1/2$, $\hat{W}_\tau/\tau$   converges as    in a binomial distribution.
This is the normal phase. 
It is the super-normal phase transition.

\end{document}